\documentclass[preprint, 3p]{elsarticle}
\usepackage{amsthm}
\usepackage{amssymb}
\usepackage{dsfont}
\usepackage{graphicx}
\usepackage{subfigure}
\usepackage{color}
\biboptions{square}
\journal{Physica A}

\begin{document}

\begin{frontmatter}

\title{\bf Dynamical generalized Hurst exponent as a tool to monitor unstable periods in financial time series }
\bigskip
\author[kcl]{Raffaello Morales\corref{cor1}}
\ead{raffaello.morales@kcl.ac.uk}
\author[kcl]{T. Di Matteo}
\author[kcl]{Ruggero Gramatica}
\author[kent,anu]{Tomaso Aste}
\address[kcl]{Department of Mathematics, King's College London, Strand WC2R 2LS, London, UK.}
\address[kent]{School of Physical Sciences, University of Kent, Canterbury, Kent, CT2 7NH, UK.}
\address[anu]{Applied Mathematics, Research School of Physics and Engineering, The Australian National University, Canberra ACT 0200, Australia.}
\cortext[cor1]{Corresponding author Tel. +44 20 78481197}

\smallskip

\begin{abstract}
We investigate the use of the Hurst exponent, dynamically computed over a moving time-window, to evaluate the level of stability/instability of financial firms. 
Financial firms bailed-out as a consequence of the 2007-2010 credit crisis show a neat increase with time of the generalized Hurst exponent in the period preceding the unfolding of the crisis. 
Conversely, firms belonging to other market sectors, which suffered the least throughout the crisis, show opposite behaviors. 
These findings suggest the possibility of using the scaling behavior as a tool to track the level of stability of a firm.   
In this paper, we introduce a method to compute the generalized Hurst exponent which assigns larger weights to more recent events with respect to older ones. 
In this way large fluctuations in the remote past are less likely to influence the recent past. 
We also investigate the scaling associated with the tails of the log-returns distributions  and compare this scaling with the scaling associated with the Hurst exponent, observing that the processes underlying the price dynamics of these firms are truly multi-scaling. 

\end{abstract}
\end{frontmatter}
\bigskip
\section{Introduction}
The search for scaling behaviors in financial markets is nowadays a very rich discipline 
\cite{mandelbrot1963variation,mandelbrot1997fractals,calvet2002multifractality,bouchaud2000apparent,mantegna1995scaling,lebaron2001stochastic, kaizoji2003scaling,scalas1998scaling,Bartolozzi07,Liu07,Liu08} 
where the growing amount of empirical data is continuously advancing the understanding of markets behaviors. 
Two types of scaling \cite{kantelhardt2002multifractal,groenendijk1998hybrid} are observed and studied in the finance literature: the first one is associated with any volatility measure and its scaling in time (e.g. moments of the returns distribution), while the second one reflects the behavior of the tails of the distribution of returns. 
In this paper we look at both of them and at the relationship between the two by using the generalized Hurst exponent (GHE) approach. 
Previous works \cite{di2007multi,matteo2005long} have highlighted that the value of the GHE allows to characterize the stage of development of a market, with values of the GHE greater than $0.5$ indicating a low stage of development, typical of the emerging markets, while values of the GHE  lower than $0.5$ corresponding to an advanced stage of development. 
Here we study whether the same paradigm can be applied to characterize the level of stability of a firm.
To this purpose we introduce a weighted average to compute the dynamical generalized Hurst exponent obtaining a finer differentiation in the historical time series by smoothing the propagation of large fluctuations from the remote past to the near present. 
Although multi-scaling analysis based on the GHE has been already extensively pursued in the literature \cite{Bartolozzi07,Liu07,Liu08,di2007multi,matteo2005long,di2003scaling, barunik2010hurst, mandelbrot1997multifractal, luxmultifractal, carbone2007algorithm}, the dynamics of the GHE has been scarcely investigated \cite{carbone2004time}. 
In this work we have used a moving time-window and studied the behavior in time of the GHE of different financial time series with the aim of both uncovering the statistical properties of the empirical data and pointing out further potentials for the applications of this tool. 
In particular, our analyses have been focused in determining whether the GHE may be used to track the stability of firms from several market sectors. 
The data are from 395 stock prices of companies listed in the New York Stock Exchange (NYSE) and have been provided by Reuters. We have analyzed several companies belonging to different market sectors but we have focused our attention on the companies most severely involved in the unfolding of the 2007-2010 ``credit crunch'' crisis. 
The scaling analysis based on the estimation of the GHE is also compared to the one associated with the behavior of the tails of the distribution. 
\newline This paper is structured as follows: section 2 recalls the definition of the GHE; section 3 describes the weighted-average algorithm; in section 4 the empirical analysis is performed and a proper choice of the parameters of the system is discussed; section 5 introduces the scaling of the distributions of the returns whose relations to the GHE is reported in section 6; conclusions are drawn in section 7. 

\section{Generalized Hurst Exponent}
The generalized Hurst exponent is a tool to study directly the scaling properties of the data via the qth-order moments of the distribution of the increments and it is associated with the long-term statistical dependence of a certain time series $S(t)$, with $t=(1, 2,\dots, k,\dots, \Delta t)$, defined over a time-window $\Delta t$ with unitary time-steps.\footnote{Here, to simplify notation, we use unitary time-steps; generalization to arbitrary time-steps is straightforward.} 
Being a measure of correlation persistence, it is necessarily related to fundamental statistical quantities which turn out to be the qth-order moments of the distribution of the increments, defined as \cite{di2007multi, barabasi1991multifractality}
\begin{equation}\label{eq1}
K_{q}(\tau)=\frac{\langle |S(t+\tau)-S(t)|^{q}\rangle}{\langle|S(t)|^{q}\rangle},
\end{equation}
where $\tau$ can vary between 1 and $\tau_{max}$ and $\langle\cdot\rangle$ denotes the sample average over the time-window. 
Note that for $q=2$, $K_{q}(\tau)$ is proportional to the autocorrelation function: $C(t,\tau)=\langle S(t+\tau)S(t)\rangle$.
The generalized Hurst exponent is then defined from the scaling behavior of $K_{q}(\tau)$ when the following relation holds:
\begin{equation}\label{eq2}
K_{q}(\tau) \propto \tau^{qH(q)} \;\;.
\end{equation}
Processes exhibiting this scaling behavior can be divided into two classes: 
(i) Processes with $H(q)=H$, i.e. independent of $q$. These processes are uniscaling (or unifractal) and their scaling behavior is uniquely determined by the constant $H$ (Hurst exponent or self-affine index \cite{di2007multi});
(ii) Processes with $H(q)$ not constant. These processes are called multiscaling (or multifractal) and each moment scales with a different exponent. 
Previous works have pointed out how financial time series exhibit scaling behaviors which are not simply fractal, rather multi-fractal, or multiscaling \cite{di2007multi,matteo2005long}. 
The GHE is computed from an average over a set of values corresponding to different values of $\tau_{max}$ in Eq.$\ref{eq1}$  \cite{matteo2005long,di2003scaling,GeneralizedHurst}.
The analysis based on the generalized Hurst exponent is very simple as all the information about the scaling properties of a signal is enclosed in the scaling exponent $H(q)$.

\section{Weighted exponential smoothing}
\begin{figure}[t!]
\centering
\includegraphics[width= 0.6\columnwidth]{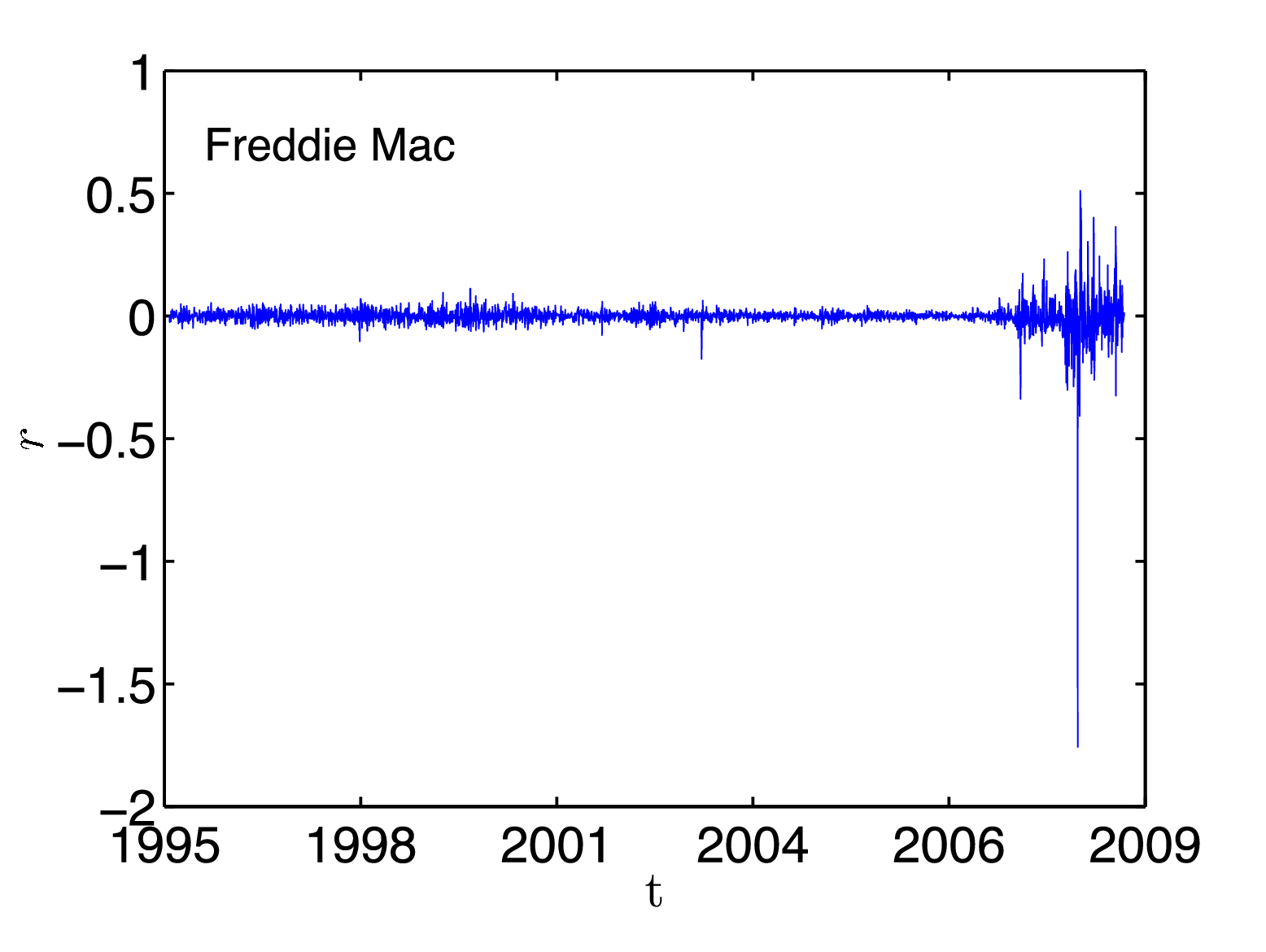}
\label{Figure1}
\caption{\textit{
Logarithmic returns, $r$, for the time series of the Freddie Mac stock prices as function of time $t$ in the period between 1 January 1996 and 30 April 2009.
The large fluctuations corresponding to the unfolding of the 2008-2009 crisis are clearly visible.}}
\end{figure}
\begin{figure}[t!]
\centering
\includegraphics[width= 0.6\columnwidth]{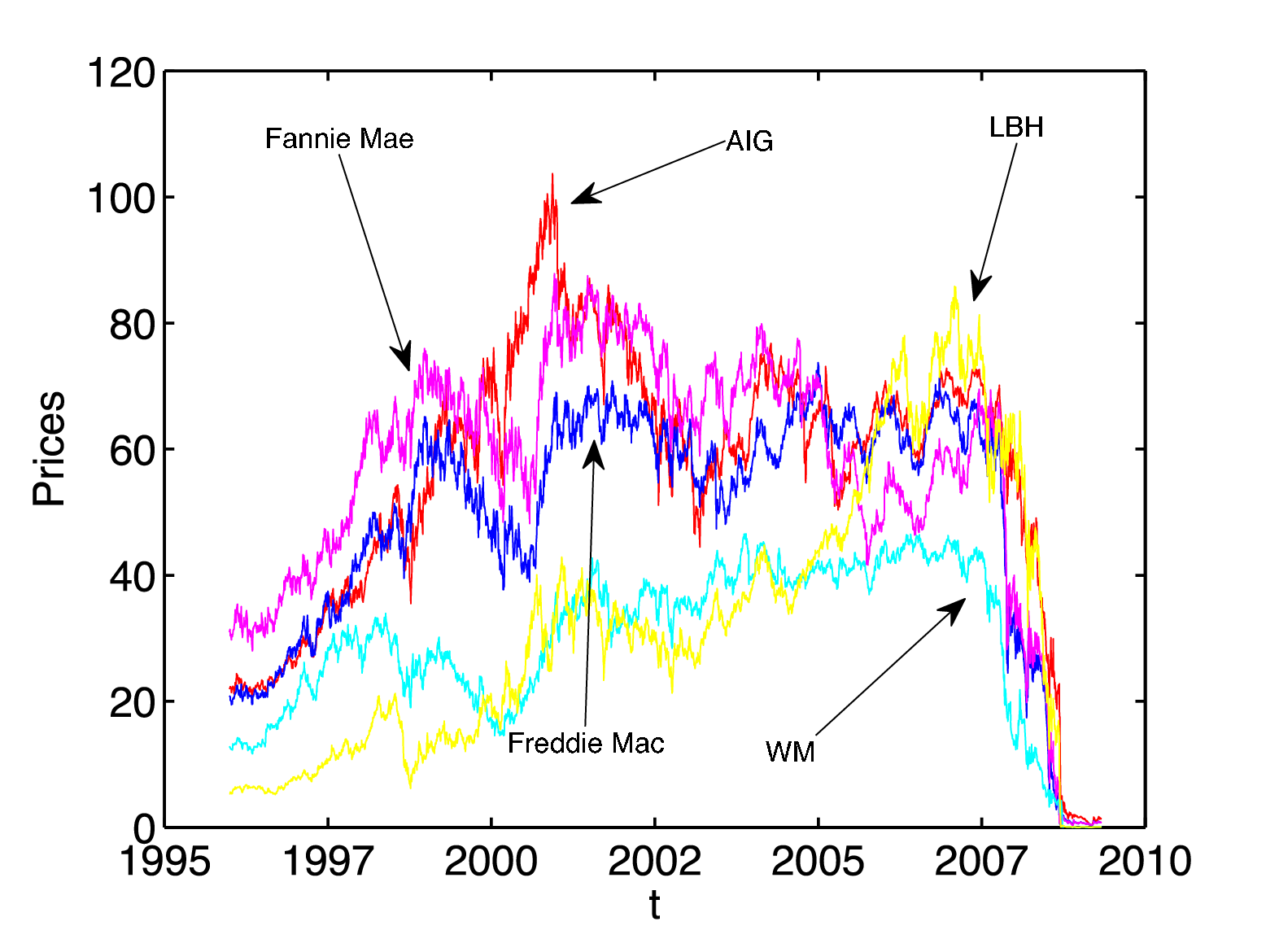}
\caption{\textit{Prices of the four bailed-out companies Fannie Mae, American International Group (AIG), Lehman Brothers (LBH), Freddie Mac, Washington Mutual (WM), as function of time.} }
\label{Figure1.1}
\end{figure} 
To take into account the fact that the recent past is more important than the remote past we can assume that the informational relevance of observations decays exponentially. This `exponential smoothing' is attained by defining weights as
\begin{equation}\label{eq3}
w_{s}=w_{0}\, \exp\left(-\frac{s}{\theta}\right), \,\, \forall \, s \in \{0,1,2,\dots, \Delta t -1 \}
\end{equation}
where $\theta$ is the weights' characteristic time and its inverse is the exponential decay factor $\alpha=\frac{1}{\theta}$. 
The parameter $w_{0}$ is given by \cite{pozzi2011exponential}
\begin{equation}\label{eq7}
w_{0}(\alpha)=\frac{1-e^{-\alpha}}{1-e^{-\alpha \Delta t}}.
\end{equation}
 The weighted average over the time-window $[t-\Delta t+1,t]$  for a general quantity $f(x_{t})$ is thus 
\begin{equation}\label{eq8}
\langle f\rangle_{w}(t)= \sum_{s=0}^{\Delta t -1} w_{s}\,f(x_{t-s})
\end{equation}
and the weighted GHE (wGHE) is therefore obtained by substituting the normal averages in Eq.$\ref{eq1}$ with weighted averages:
\begin{equation}
K_{q}^{w}(t,\tau)=\frac{\langle |S(t+\tau)-S(t)|^{q}\rangle_{w}(t)}{\langle|S(t)|^{q}\rangle_{w}(t)}.
\end{equation}
From the scaling law in Eq.$\ref{eq2}$ this leads to the linear relation
\begin{equation}\label{Hw(1)}
\ln (K_{q}^{w}(t,\tau)) = qH^{w}(q){\ln\left(\tau \right)} + const.
\end{equation}
from which the wGHE can be computed.
In the next section we apply this tool to the empirical time series. 

\section{Empirical Analysis}
\begin{figure}[h!]
\centering
\mbox{\subfigure{\includegraphics[width= 0.5\columnwidth]{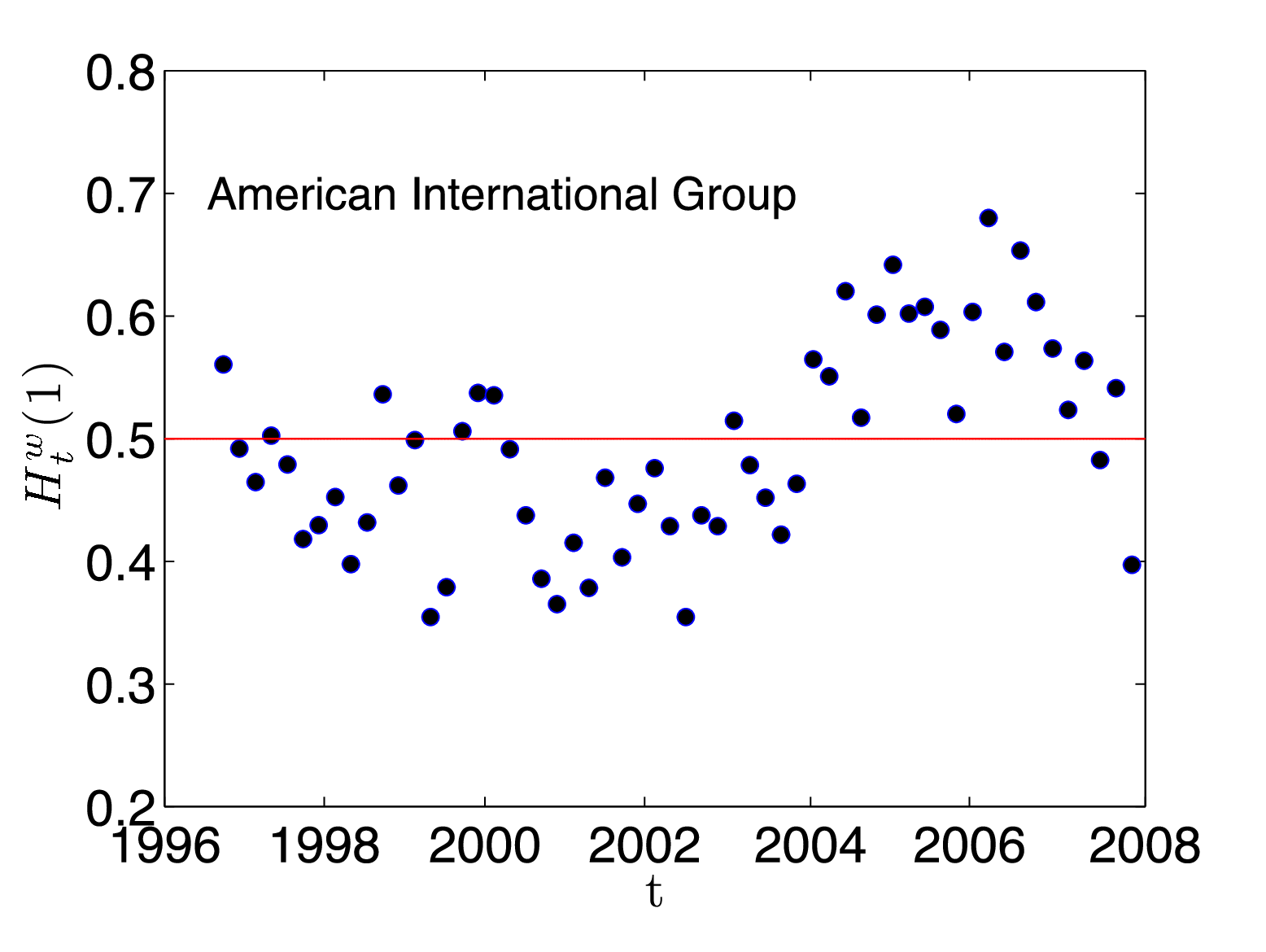}}\quad
\subfigure{\includegraphics[width= 0.5\columnwidth]{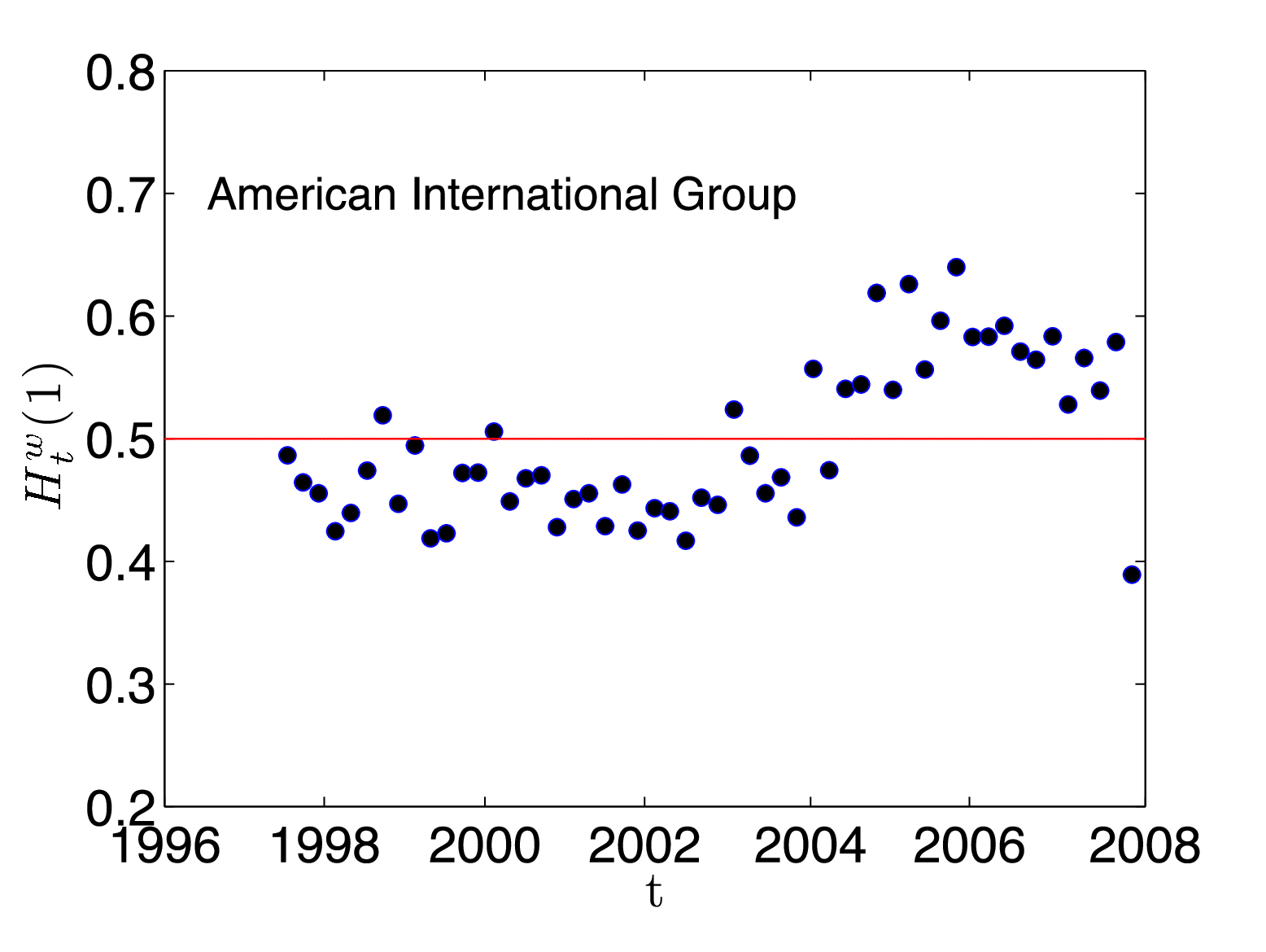} }}
\caption{
\textit{
Weighted Generalized Hurst exponent $H^w(q=1)$ as a function of time for American International Group (AIG). 
Left panel: $\Delta t = 200$ days time-window. 
Right panel: $\Delta t = 400$ days time-window. 
The characteristic time is kept constant at $\theta=300$ days in both plots.
The points are reported in correspondence of the end of the time-window.
}
}
\label{Figure2}
\end{figure}

\begin{figure}[h!]
\centering
\mbox{\subfigure{\includegraphics[width= 0.5\columnwidth]{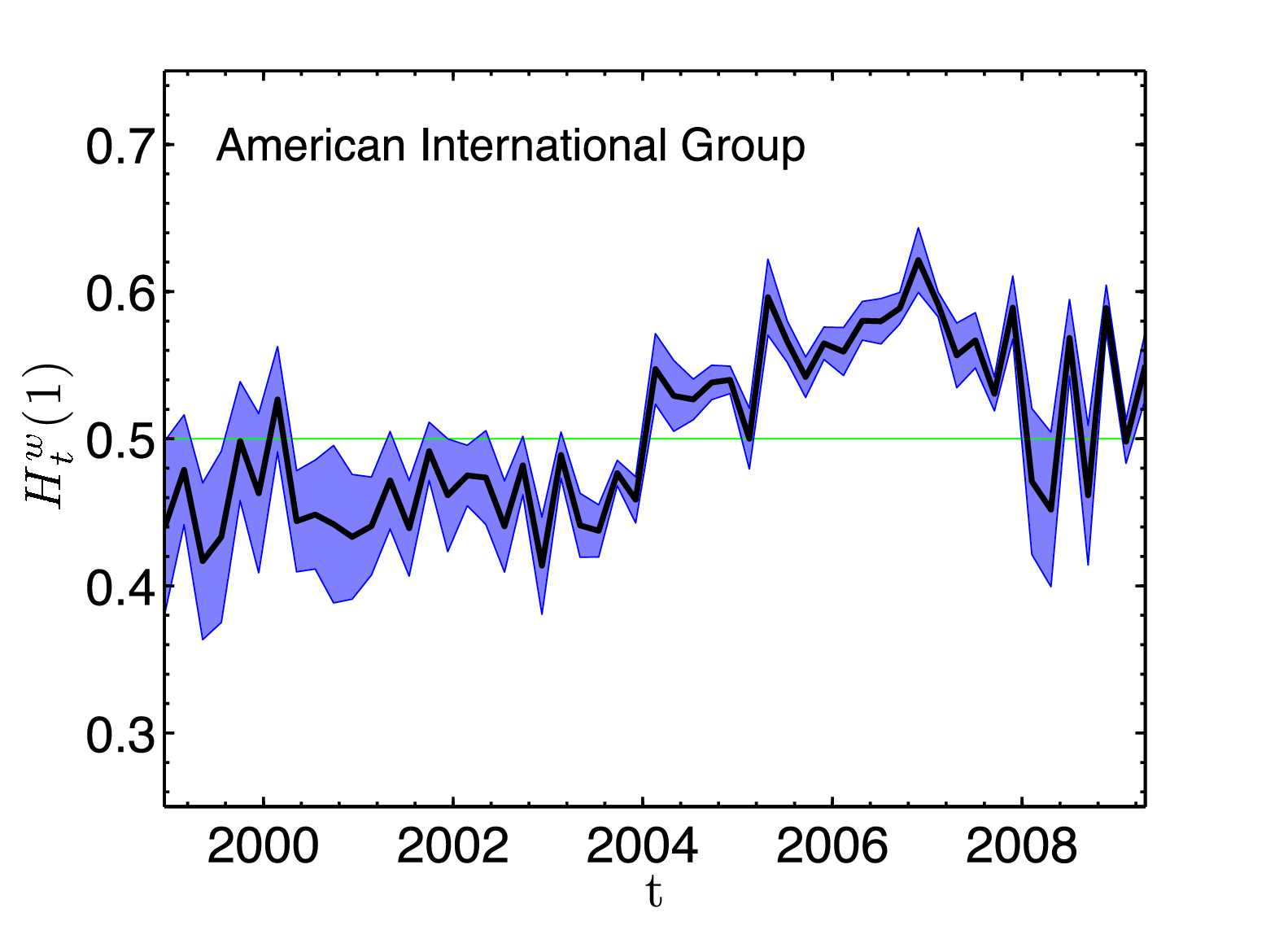}}\quad
\subfigure{\includegraphics[width= 0.5\columnwidth]{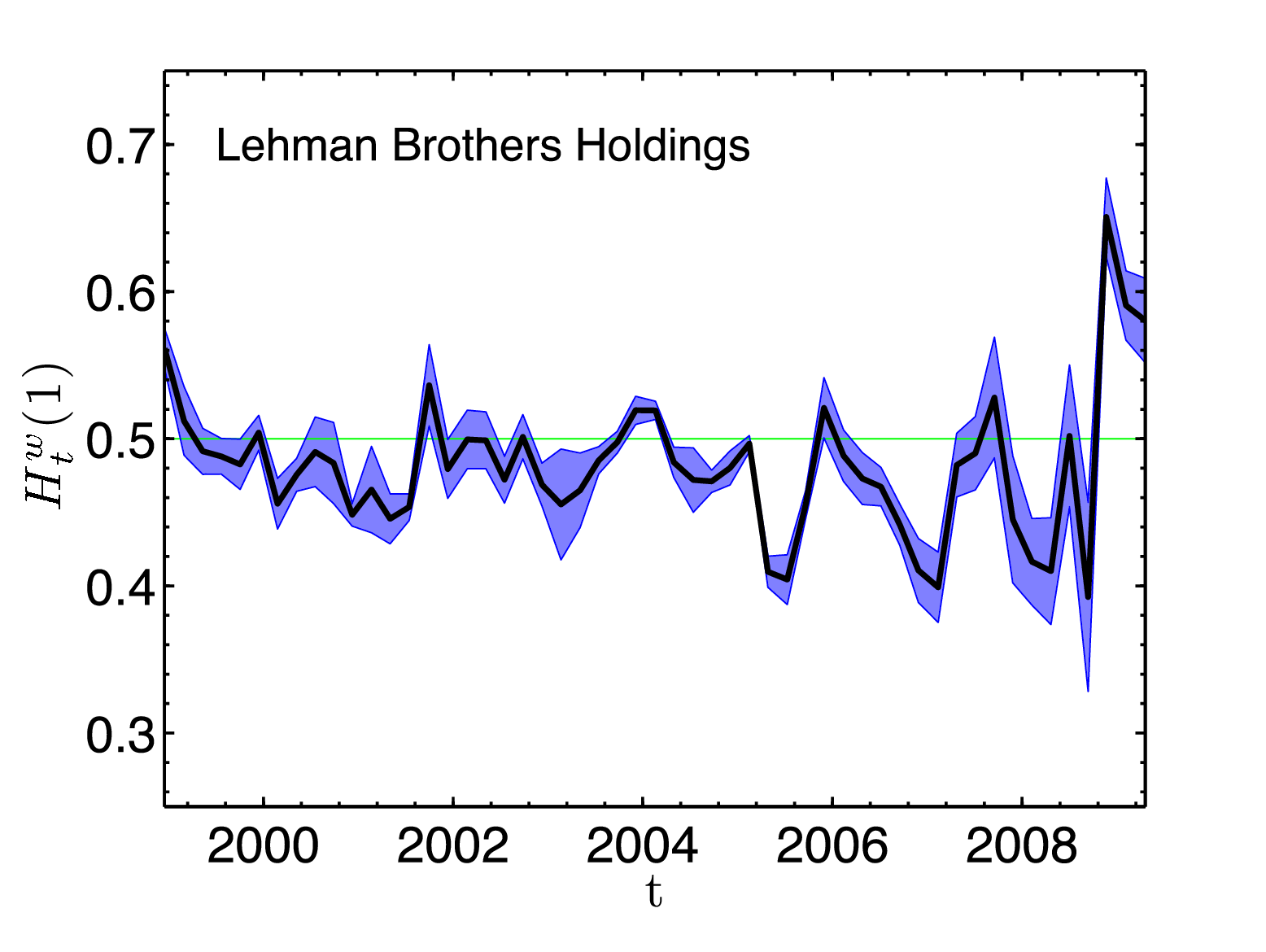} }}
\caption{\textit{ 
Left panel: weighted Generalized Hurst exponent as a function of time for American International Group (AIG). 
Right panel: weighted Generalized Hurst exponent as a function of time for Lehman Brothers Holdings (LBH). 
Note the abrupt jump in the value of the GHE at the end of the time-period. 
The overlapping time-windows are $\Delta t = 750$ days, with $\theta=250$ days. 
The values are plotted in correspondence of the end of the time-window (solid black line).
The shaded areas around the tick-line plot represent the sizes of the standard deviations. 
}
}
\label{Figure3}
\end{figure}

\begin{figure}[h!]
\centering
\mbox{\subfigure{\includegraphics[width= 0.5\columnwidth]{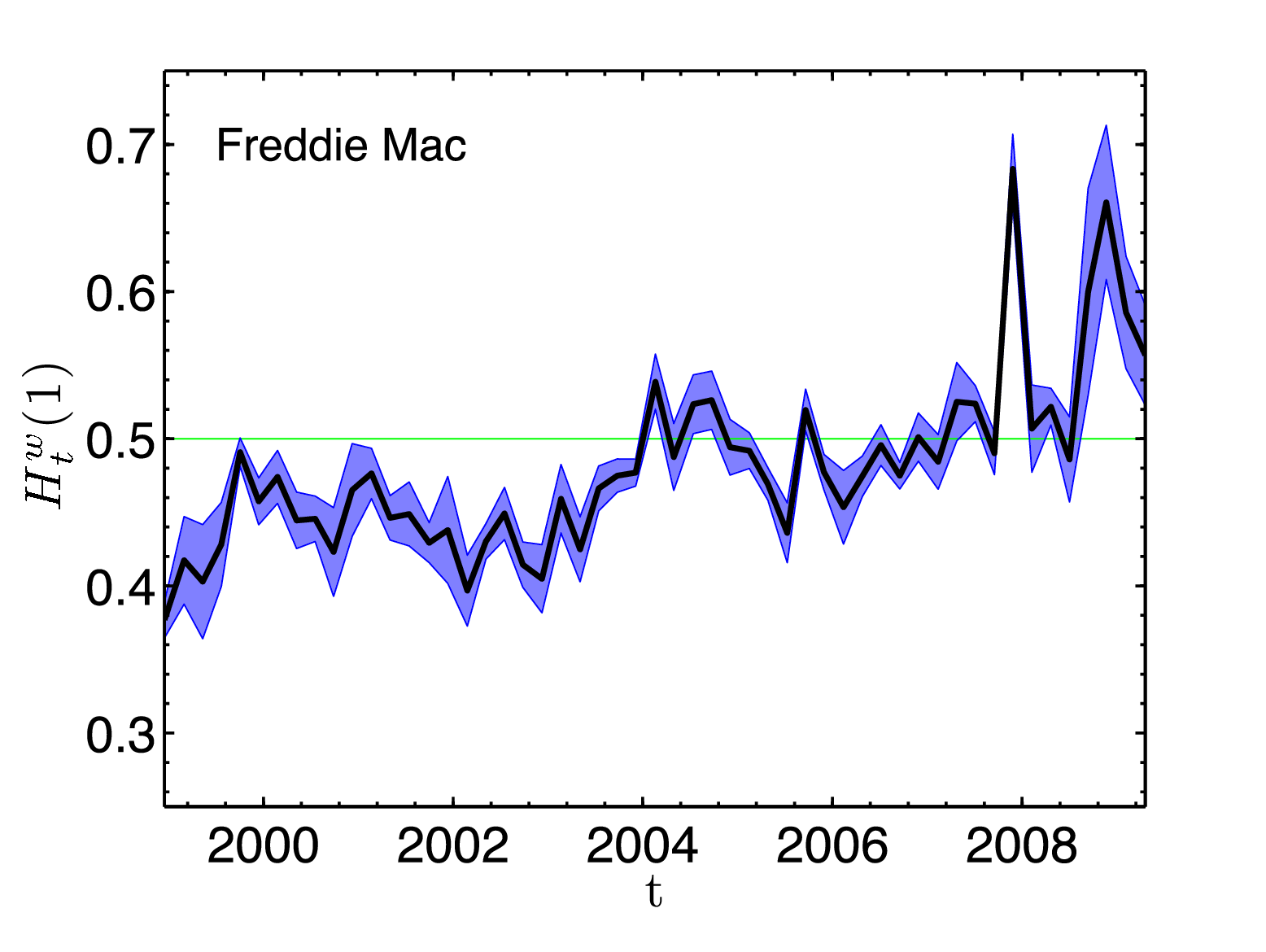}}\quad
\subfigure{\includegraphics[width= 0.5\columnwidth]{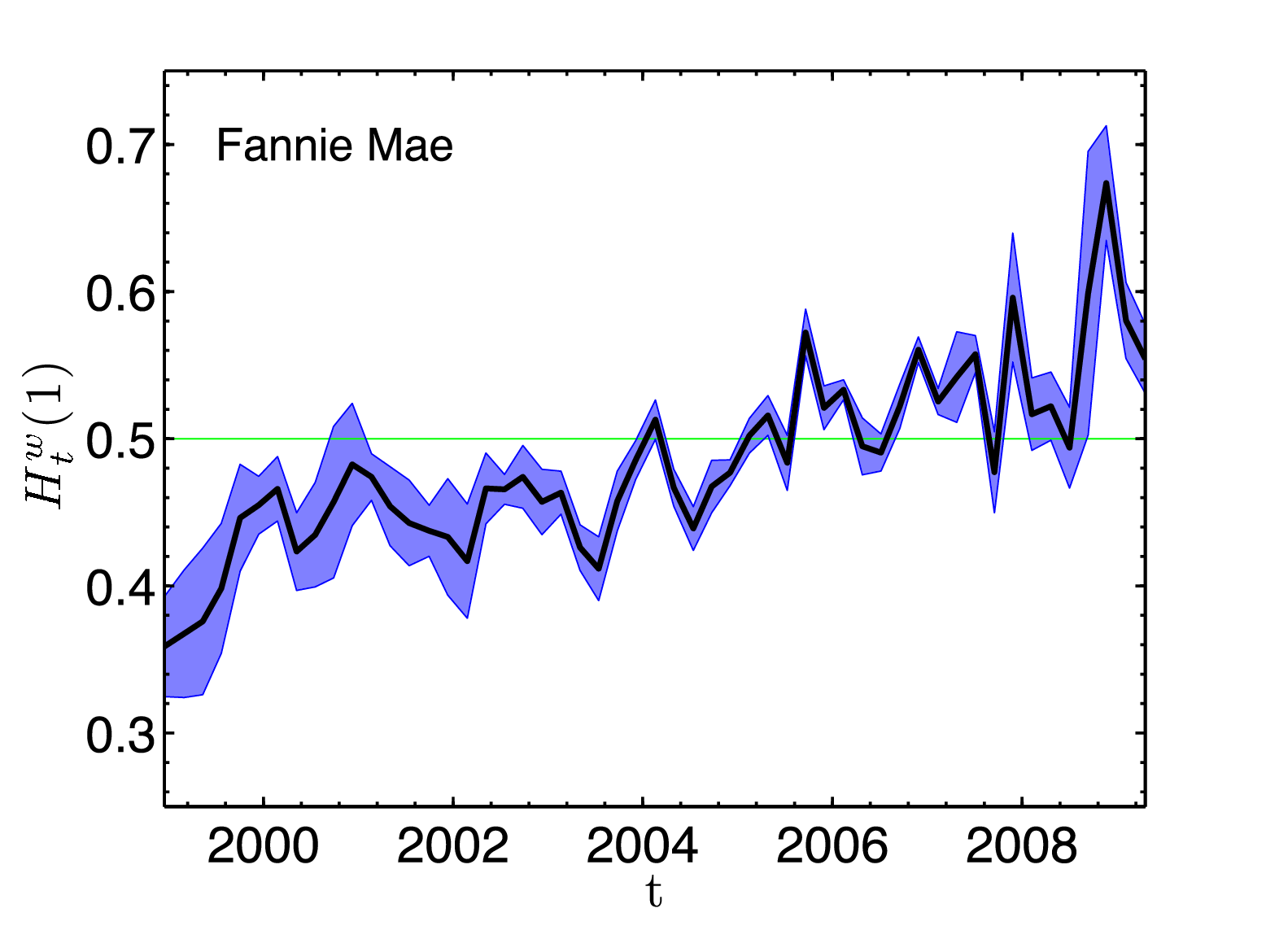} }}
\caption{\textit{
Weighted Generalized Hurst exponent $H^w(q=1)$ as a function of time. 
Left panel: Freddie Mac. Right panel: Fannie Mae. 
The increasing trend over the whole period highlights a transition from values of $H^w(1)<0.5$ to values of $H^w(1)>0.5$. 
This suggests a progressive change in the stability of the companies under study.
} 
}
\label{Figure4}
\end{figure}

\begin{figure}[h!]
\centering
\mbox{\subfigure{\includegraphics[width= 0.5\columnwidth]{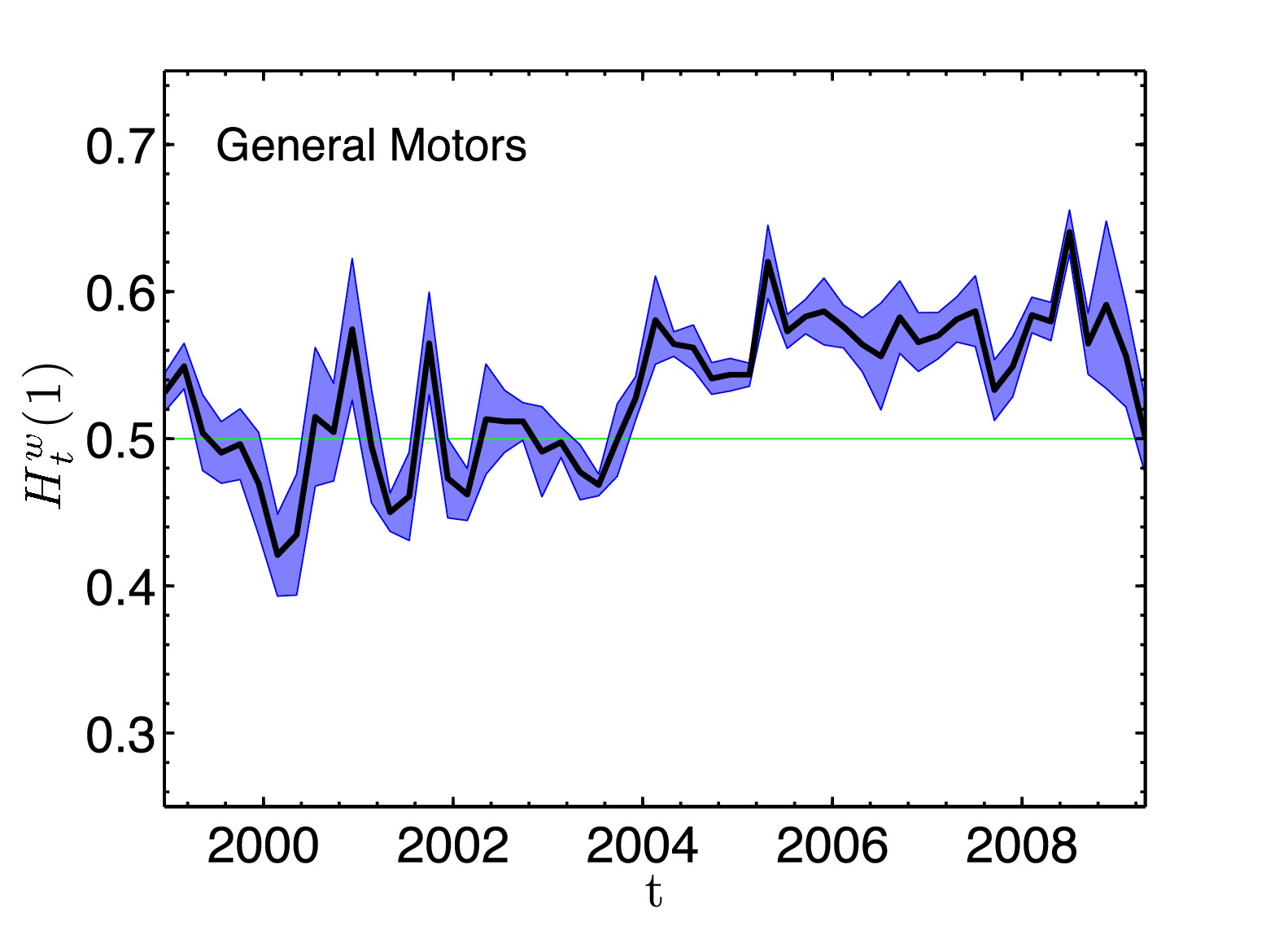}}\quad
\subfigure{\includegraphics[width= 0.5\columnwidth]{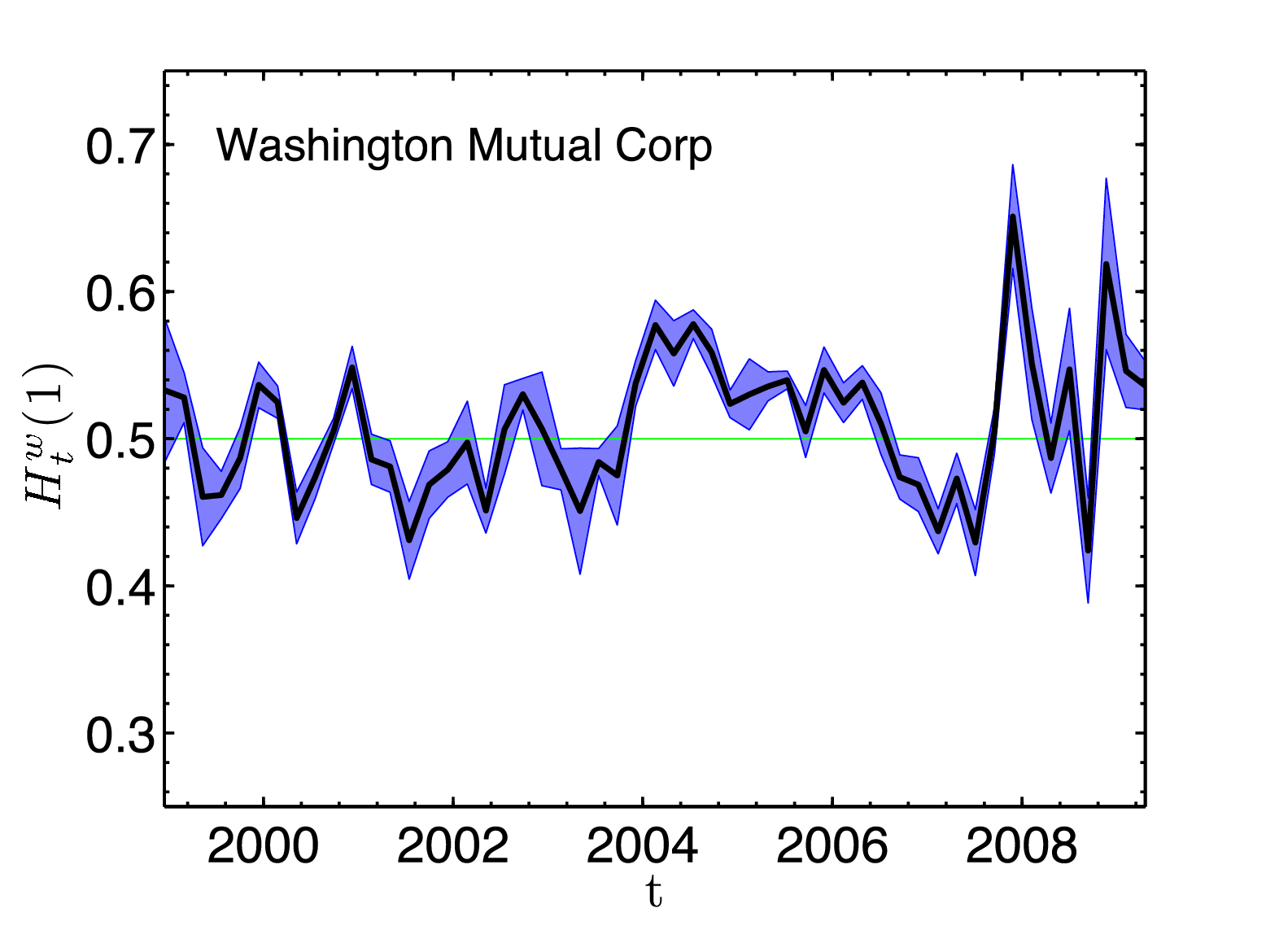} }}
\caption{\textit{
Weighed Generalized Hurst exponent $H^w(q=1$) as a function of time. 
Left panel: General Motors, a company that went bankrupt following Chrysler in June 2009. 
Its bankruptcy was classified as the fourth largest in U.S. history. 
Right panel: Washington Mutual. 
The increasing trend over the whole period highlights a transition from values of $H^w(1)<0.5$ to values of $H^w(1)>0.5$. 
This suggests a progressive change in the stability of the companies under study. 
} 
}
\label{Figure5}
\end{figure}

\begin{figure}[h!]
\centering
\mbox{\subfigure{\includegraphics[width=0.5\columnwidth]{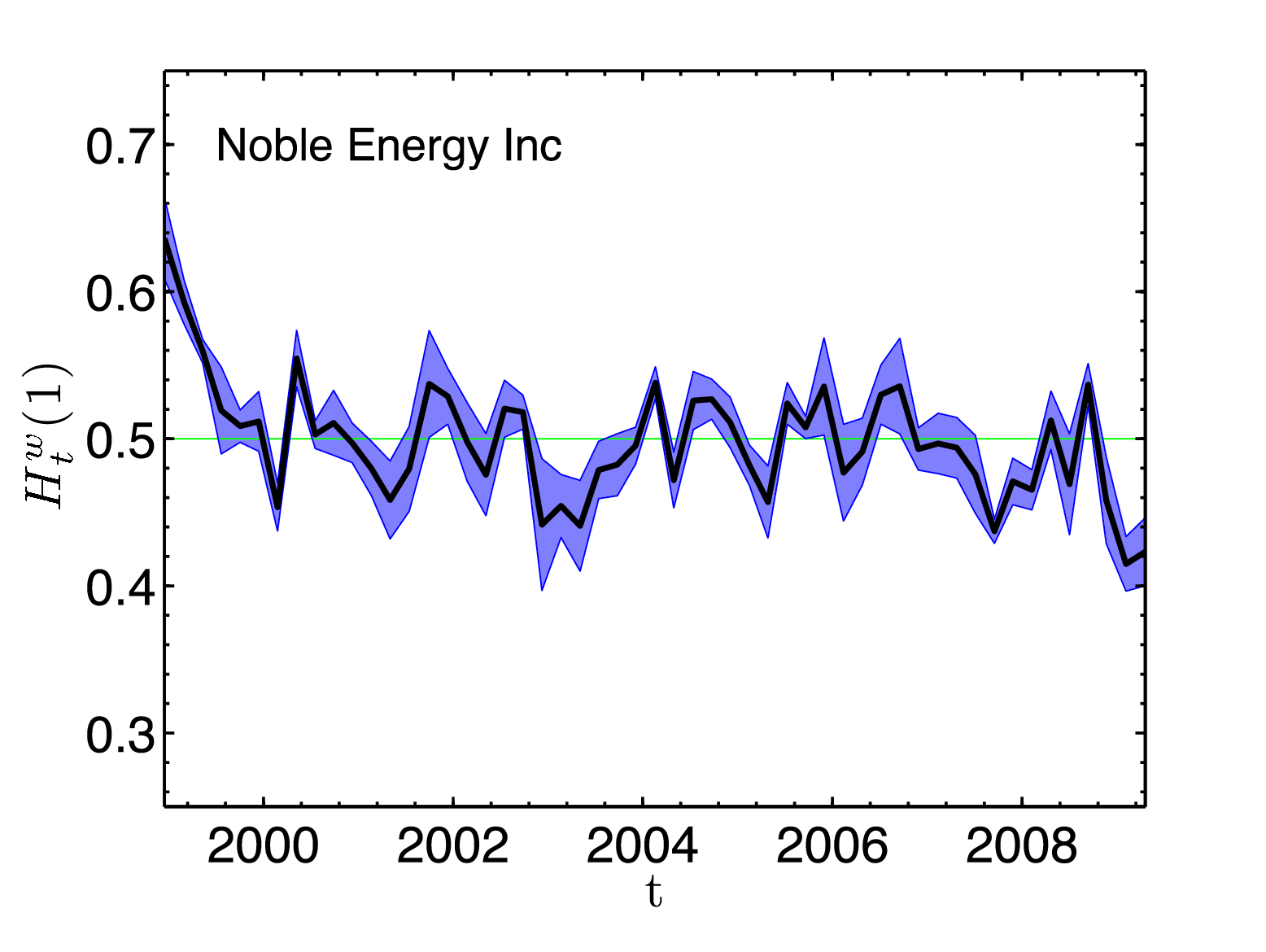}}\quad
\subfigure{\includegraphics[width= 0.5\columnwidth]{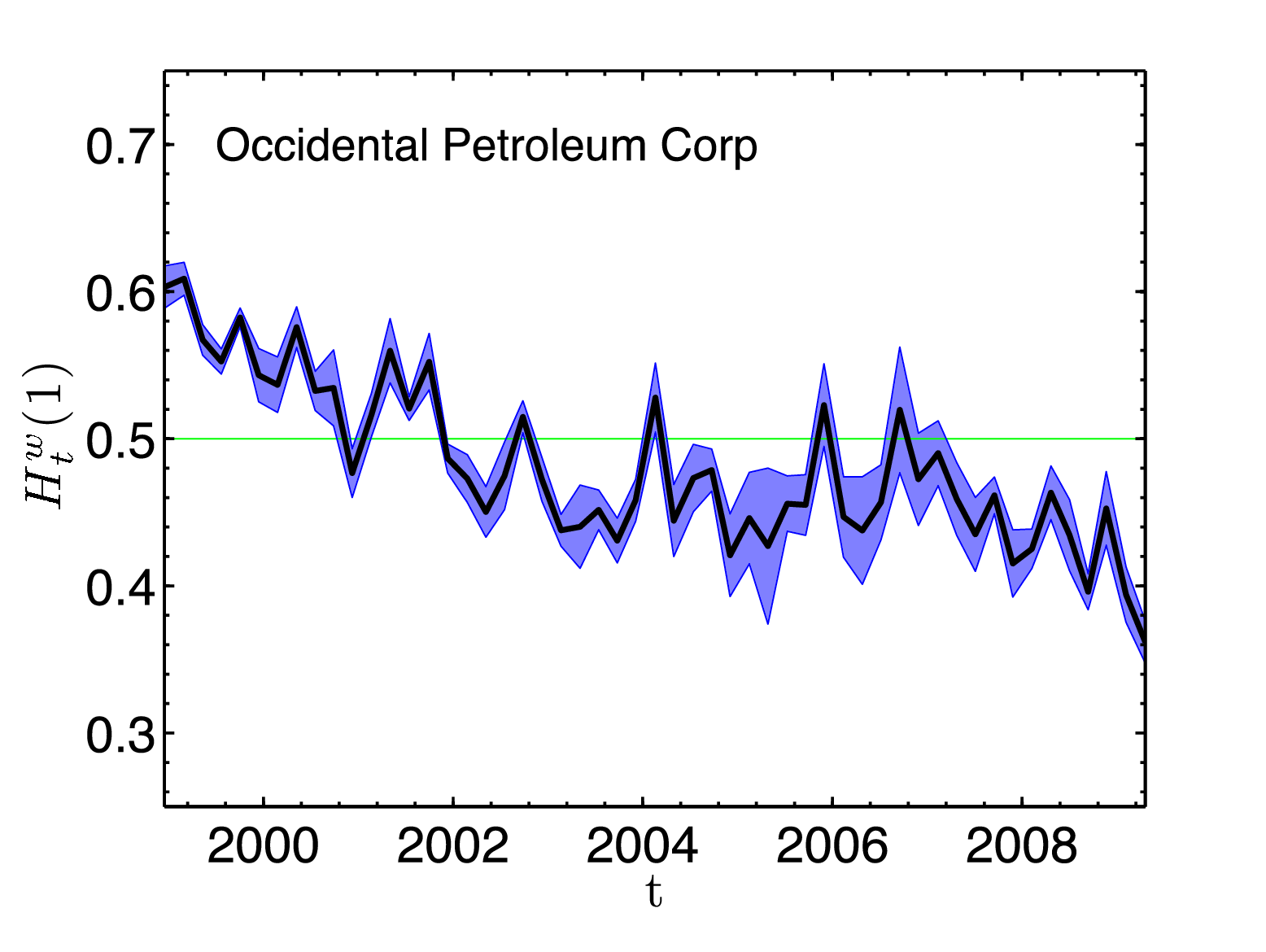} }}
\caption{\textit{
Weighed Generalized Hurst exponent $H^w(q=1$) as a function of time for: Left panel - Noble Energy Inc.; Right panel - Occidental Petroleum. 
The time-window is taken to be $\Delta t=750$ days and $\theta=250$ days. 
}
}
\label{Figure6}
\end{figure}

The empirical time series here analyzed include daily stock prices from 1 January 1996 to 30 April 2009. 
From these prices we define a new time series of the daily log-returns
\begin{equation}\label{eq9}
r(t)=\ln(P(t+1))-\ln(P(t))
\end{equation}
where $P(t)$ is the daily price. 
In Fig.1 an example of log-returns for the  Freddie Mac stock prices is shown.
Not surprisingly, these returns exhibit large fluctuations in the crisis period. 
From the log-returns we have then computed the wGHE by using Eq.\ref{Hw(1)}.
In analogy with \cite{di2003scaling,matteo2005long,GeneralizedHurst} we have estimated the $H^w(q)$ as an average of several linear fits of Eq.\ref{Hw(1)} with $\tau\in[1,\tau_{max}]$ and varying $\tau_{max}$ between 5 and 19.
As proxy of the statistical uncertainty of the scaling law we have computed the standard deviation of the $H^w(q)$ over this range of  $\tau_{max}$.
To track the evolution of the stage of development of a certain company, we have studied the dynamics in time of the wGHE on overlapping time-windows with a constant 50 days shift between any two successive windows. 
\\
First of all, to fully capture the advantages of the weighted average method, a choice of the parameters $\theta$ and $\Delta t$, namely the characteristic time and the width of the time-window, has to be made. 
In particular the time-window  $\Delta t$ must be large enough to provide good statistical significance but it should not be too large in order to retain sensitivity to changes in the scaling properties occurring over time. 
In order to satisfy both these requirements we take a rather long time-window $\Delta t$ combined with a relatively short characteristic time  $\theta$.
For instance, in Fig.$\ref{Figure2}$ we show how the manipulation of the parameters $\theta$ and $\Delta t$ affects the dynamics of the Hurst exponent of the company American International Group (AIG). 
As it can be appreciated in the figure, which shows plots for AIG with time-windows of respectively 200 days (left panel) and 400 days (right panel) while keeping $\theta=300$ days, the shape of the outline shrinks and gets neater as the time-window is increased. 
The left panel of Fig.$\ref{Figure2}$ shows more noisy dynamics when $\Delta t$ is smaller. 
Conversely, in the right panel we can appreciate that a slimmer outline is achieved by increasing the statistics, but duly weighting it. 
This can be further improved by increasing the value of $\Delta t$ up to three years of trading time while keeping $\theta$ down to one year.
The result of this is shown in Fig.$\ref{Figure3}$ for Lehman Brothers Holdings (LBH) and American International Group  where the thick lines are the average $H^w(1)$ and the bands are given by the standard deviations over  $\tau_{max}$ between 5 and 19 days \cite{di2003scaling,matteo2005long,GeneralizedHurst}.
This choice of the parameters is probably the best as it allows to have a sufficiently large, though not too much, statistics, but at the same time the events are weighted such that not to all the information present in the time series is given the same importance. 
\newline Once the choice of the parameters is made, one can notice that the behavior of $H^w(1)$ for AIG is slightly different from the one of LBH. The first one shows indeed a well-defined increasing trend, with a transition from values $<0.5$ to values $>0.5$, while LBH keeps steady around $0.5$, except for a decrease towards the end of the period followed by a sudden leap upwards. 
\newline In Fig.$\ref{Figure4}$ the dynamics in time of $H^w(1)$ for the companies Freddie Mac and Fannie Mac is reported. These are public government sponsored enterprises which in September 2008 had to be put into conservatorship by the U.S. Treasury; namely the huge debts of these companies were purchased by the U.S. government. After playing a central role in the market during the mortgages's boost both firms defaulted. Their fate is pretty well pictured by the dynamical wGHE.
Indeed, there is a clearly visible trend in these plots showing how the value of $H^w(1)$ for these companies has been increasing since 1996 until 2009. 
This is particularly interesting if we compare the two panels. According to \cite{di2007multi,matteo2005long} these trends might suggest a transition from a stable stage of the companies to an unstable one. 
\newline Other bailed-out companies which show the same trend are shown in Fig.$\ref{Figure5}$. Again the trend is increasing and crossing over the value of $0.5$ towards the end of the time-period when the crisis fully unfolded. 
\newline We have compared these results with those obtained by looking at  other companies either from the financial sector or belonging to other market sectors to test the significance of these results. 
For example, in the Basic Materials sector, we find many companies whose dynamical wGHE decreases in time, thus exhibiting an opposite behavior to that shown by the bailed-out companies in the financial sector. 
An example is reported in Fig.$\ref{Figure6}$ where the dynamical wGHE's for two companies belonging to the sector of Basic Materials are shown. 
We notice a very definite overall decreasing trend, as if the companies securities gained persistence in going through the period of crisis. 
This is in agreement with what has been considered as the boost of the commodities market during the crisis, where investors were turning away from the financial sector. 
\newline There are other sectors that  have revealed instead no particular trend in the dynamical wGHE. We stress that even in the Financial sector itself, the increasing trend found for the bailed-out companies is not common to others; for instance, many companies, like American Express Co and Morgan Stanley show stable behaviors, with wGHE values steadily fluctuating about 0.5. 
We will see in the next paragraph that the sectors exhibiting a defining trend in the dynamical wGHE are also those showing extreme values in the tail exponents of their distributions of returns. 
Although the increase or decrease of the wGHE is not simply related with the return statistics only, both behaviors are associated with the fluctuations of the log-returns distributions.
\section{Fat-tails and extreme events}
\begin{figure}[h!]
\centering
\mbox{\subfigure{\includegraphics[width= 0.5\columnwidth]{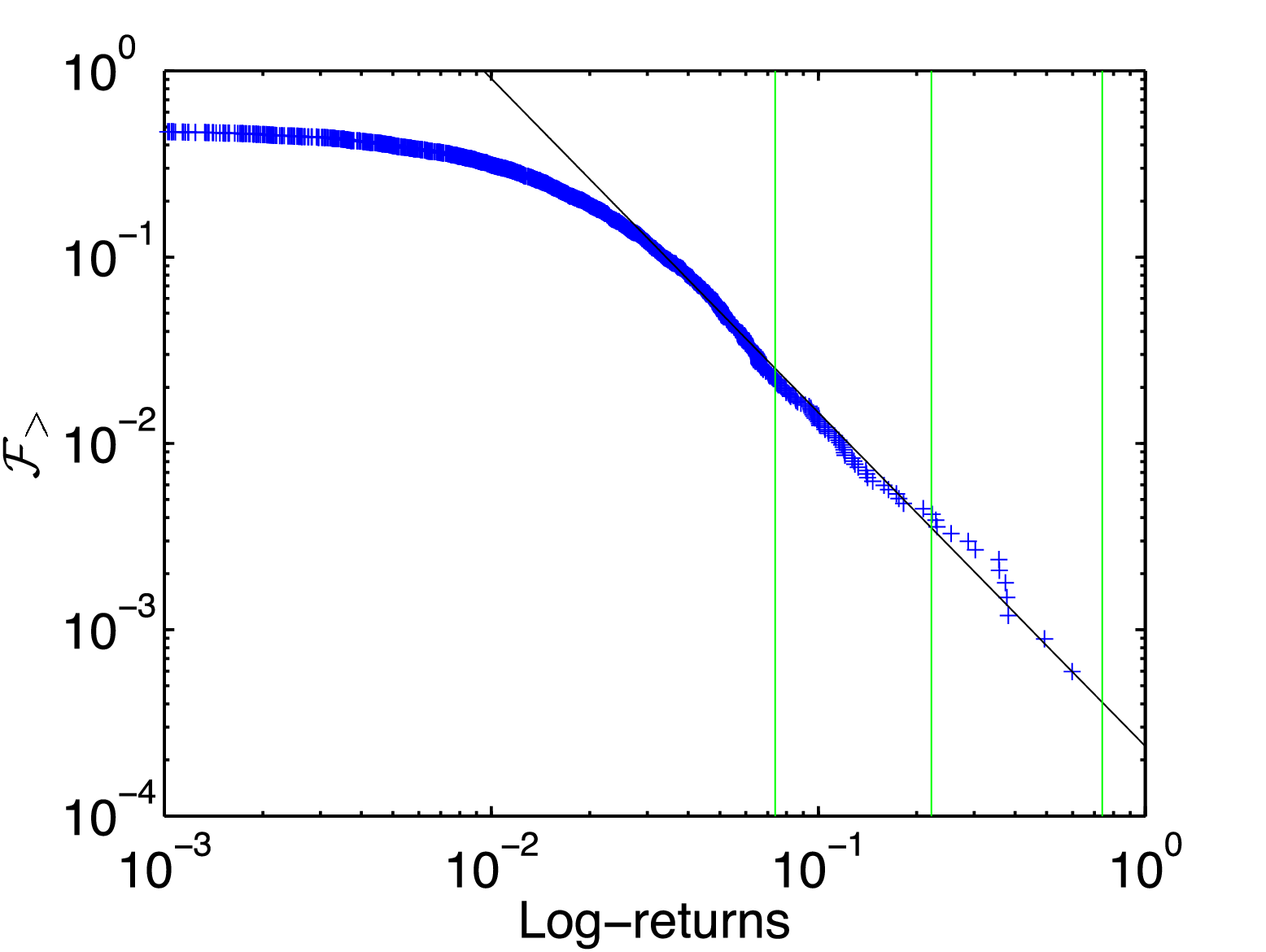}}\quad
\subfigure{\includegraphics[width= 0.5\columnwidth]{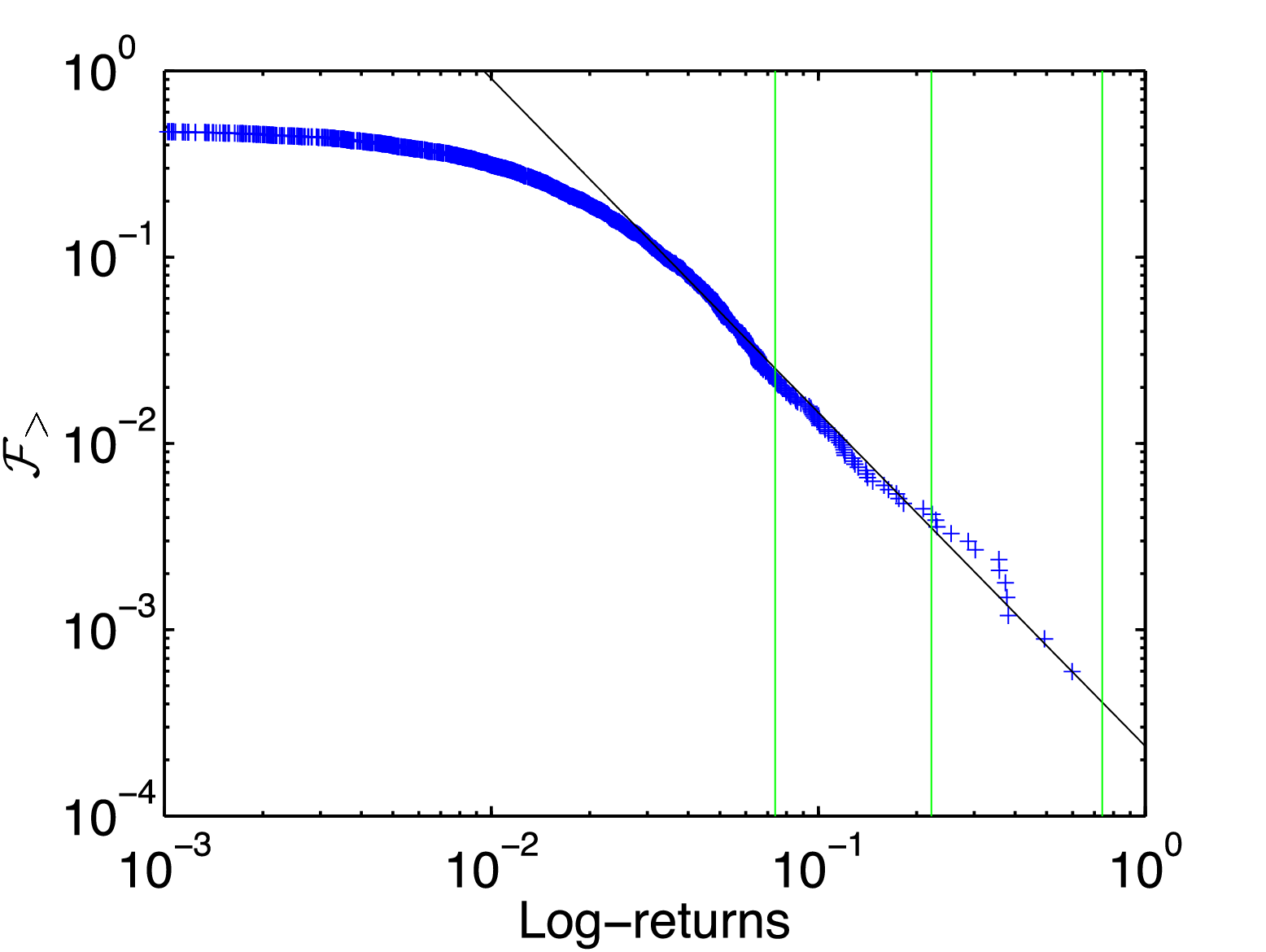} }}
\caption{\textit{
Complementary cumulative distributions of the log-returns for the stock prices of Lehman Brothers (left panel) and American International Group (right panel). 
The vertical green lines mark respectively one, three and ten standard deviations (from left to right). 
The black line is the best fit of the tail region with the power law function $\mathcal{F}_{>}(r)\propto r^{-\alpha}$. 
The estimated best-fit exponent is $\alpha \sim 1.7$ for both companies.
}}
\label{Figure7}
\end{figure}

\begin{figure}[h!]
\centering
\includegraphics[width=0.7\columnwidth]{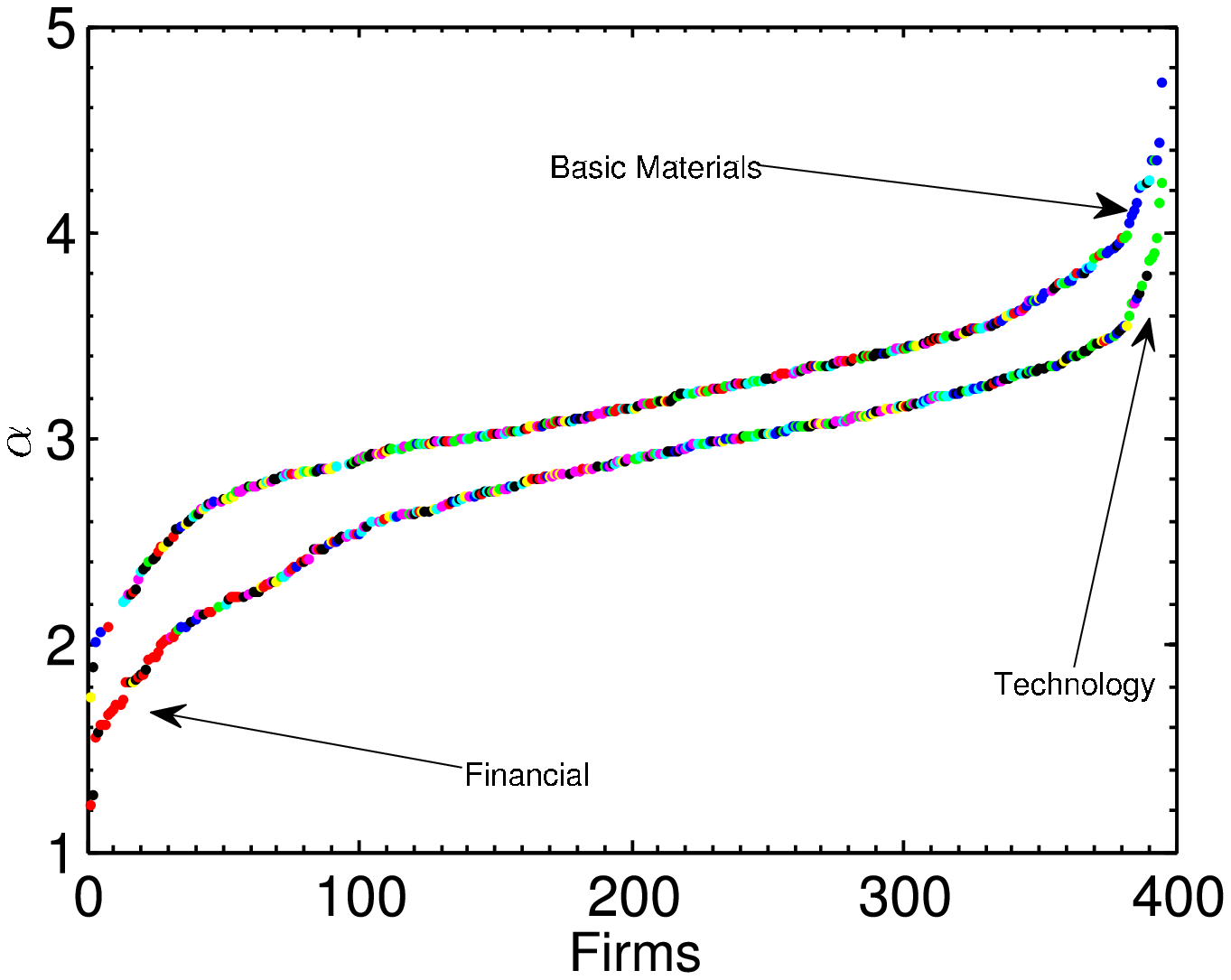}
\caption{\textit{
(Color Online) The tail exponents for all the companies analyzed including (lower curve) and excluding (upper curve) the time-period from December 2007 to April 2009, when the crisis occurred.  
We notice a clustering of the financial sector (red) at very low values of $\alpha$, with many points lying in the region $\alpha<2$. 
The other end of the curve, at high values of $\alpha$, is instead mostly populated by the Technology (green), which has been the less affected by the crisis. 
This is in agreement with the renown fact that the financial sector was the one most profoundly affected by the crisis and whose fluctuations were the largest. 
Instead, before the crisis, the sector of Basic Material (blue) appears to be the most stable.
}}
\label{Figure8}
\end{figure}

The unfolding of the 2008-2009 "credit crunch'' financial crisis has made all of us again aware that in markets very large fluctuations can happen with finite probability. 
Indeed large fluctuations are very unlikely, say impossible, in a normal statistics frame but are instead rather common in complex systems and they are properly accounted by non-normal statistics. 
In order to quantitatively catch these large fluctuations we have investigated the scaling of the tails of the distributions of the log-returns. 
In Fig.$\ref{Figure7}$ we report the complementary cumulative distributions for the stock prices of the same companies studied in the previous section. Let us recall that, given a probability density function $F(x)$, its complementary cumulative distribution \footnote{The plot of $\mathcal{F}_{>}$ in Fig.$\ref{Figure7}$ is a so-called rank-frequency plot. This is a very convenient and simple method to analyze the tail region of the distribution without any loss of information which would instead derive from gathering together data points with an artificial binning. In order to make this plot from a given set of observations $\{x_{1},x_{2},\dots, x_{T}\}$, one first sorts the T observed values in ascending order and then plots them against the vector $[1,(T-1)/T, (T-2)/T, \dots, 1/T]$. Indeed, we have that $Rank(x_{i})=1-\mathcal{F}_{>}(x_{i})$} is defined by
\begin{equation}\label{eq10}
\mathcal{F}_{>}(x)=1-\mathcal{F}_{<}(x)=1-\int_{-\infty}^{x}F(s)ds.
\end{equation}
We can see from Fig.\ref{Figure7} that fluctuations above $3\sigma$ have frequencies above $10^{-2}$ and therefore are occurring on average several times a year. We can also observe that the tails decrease linearly in  log-log scale.
Indeed, we find, in the tail region, good fits with the power law function $\mathcal{F}_{>}(r) \propto r^{-\alpha}$ with $\alpha \sim 1.7$.
Although the linear decrease of large fluctuations in log-log scale is not necessarily a proof for power-law behavior, in this case the power law hypothesis is enforced by the p-value test ($p=0.43$ for AIG and $p=0.48$ for LBH) \cite{clauset2007power}. 
However we stress that by excluding the recent unstable period from the same dataset, i.e. taking off the years 2008-2009, a slightly different picture emerges with the scaling exponents exhibiting larger values and the frequency of very large fluctuations becoming an order of magnitude smaller. 
Fig.$\ref{Figure8}$ shows the exponents for all the firms, computed both over the entire period and over the period excluding the crisis. 
One can note from Fig.\ref{Figure8} that, excluding the crisis period, the exponent increases across all firms and the occurrence of extreme events is much lower than the one observed when the crisis is included. 
In particular Fig.$\ref{Figure8}$ shows how the financial sector forms a cluster at the bottom end of the sorted companies, when the crisis period is included. 
It's also interesting to note that the firms belonging to the Technology sector appear to be the most stable. 
\newline 
Values of the scaling exponents $\alpha$ between 2 and 4 are commonly observed in these systems \cite{bouchaud2003theory,mantegna2000introduction}. 
These distributions typically have finite second moment $\sigma^{2}=\langle(x-\langle x \rangle)^{2}\rangle$ but diverging larger moments and this explains in turn why we find very large values for the excess kurtosis (139 for AIG and 761 for LBH).  
The fact that the tail exponents change by including or excluding in the statistics data referring to some extreme events is not a surprise though. 
It is not a surprise either, the fact that stock prices do not obey normal statistics. 
Nonetheless these large fluctuations over the last time-period when the crisis was unfolding may be somehow the cause for the increase of the wGHE, and this is what we are going to discuss in the next section.

\section{Discussion}
In order to understand the link between the two types of scaling, let us first investigate the simple ideal case where the underlying process is a random walk with $x(t)=x(t-1)+\eta(t-1)$ where $x(t)=\ln(P(t))$. 
In this case, for an arbitrary $\tau$, the log-returns $r(t,\tau)=x(t+\tau)-x(t)$ can be written as a sum of $n=\tau$ random variables:
\begin{equation}\label{eq11}
r(t,\tau)=\sum_{s=0}^{\tau -1} \eta(s+t).
\end{equation}
If the $\eta(t)$ are iid, the Central Limit Theorem applies to the $r(t,\tau)$ and there are two cases: (1) the probability distribution function of $\eta(t)$ has finite variance and therefore the distribution of $r(t,\tau)$ converges to a normal distribution for large $\tau$;  (2) the variance is not defined and the asymptotic distribution of the $r(t,\tau)$ converges to a Levy Stable distribution. For distributions well approximated by power-law functions in the tail region, the parameter that distinguishes between these two cases is the tail index $\alpha$. Namely $\alpha \ge 2$ leads to normal distributions, while $\alpha <2$ leads to Levy Stable distributions. Moreover, given that $r(t,\tau)$ is a sum of random variables and given that both cases (1) and (2) lead to stable distributions\footnote{A distribution is stable if and only if, for any $n>1$, the distribution of $y=x_{1}+x_{2}+\dots +x_{n}$ is equal to the distribution of $n^{1/\alpha}x+d$, with $d\in \mathbb{R}$. 
This implies
\begin{equation}\label{eq12}
p_{n}(y)=\frac{1}{n^{1/\alpha}} p\left(\frac{y-d}{n^{1/\alpha}}\right)
\end{equation}
where $p_{n}(y)$ is the aggregate distribution of the sum of the i.i.d. variables and $p(x)$ is the distribution of the $x_{i}$.
}
, the probability distribution $p_{\tau}(r)$, of the log-returns must scale with $\tau$ as \cite{bouchaud2003theory,mantegna2000introduction} 
\begin{equation}\label{eq13}
p_{\tau}(r)=  \left\{
\begin{array}{ll}
\frac{1}{\tau^{1/\alpha}} p\left (\frac{r}{\tau^{1/\alpha}}\right)
& \mbox{if $\alpha < 2$}\\
\\
\frac{1}{\tau^{1/2}} p\left (\frac{r}{\tau^{1/2}}\right)
& \mbox{if $\alpha \ge 2$} \;.\\
\end{array}
\right. 
\end{equation}
Accordingly, the q-moments scale as
\begin{equation}\label{eq14} 
E(|r(t, \tau)|^{q})=\left\{
\begin{array}{ll}
\tau^{q/\alpha} E(|r(t,1)|^{q}) 
& \mbox{if $\alpha < 2$}\\
\\
\tau^{q/2} E(|r(t,1)|^{q}) 
& \mbox{if $\alpha \ge 2$} \;.\\
\end{array} \right. 
\end{equation}
Here $E(...)$ denotes the expectation value. Finally, if we restrict to the class of self-affine processes, i.e. those processes $x(t)$ where the probability distribution of $\{x(ct)\}$ is equal to the probability of $\{c^{H}x(t)\}$, for any positive $c$, and we consider stationary increments, the q-moments must scale as
\begin{equation}\label{eq15}
E(|r(t, \tau)|^{q})=c(q)\tau^{qH} \;.
\end{equation}
\begin{figure}[t!]
\centering
\mbox{\subfigure{\includegraphics[width=0.5\columnwidth]{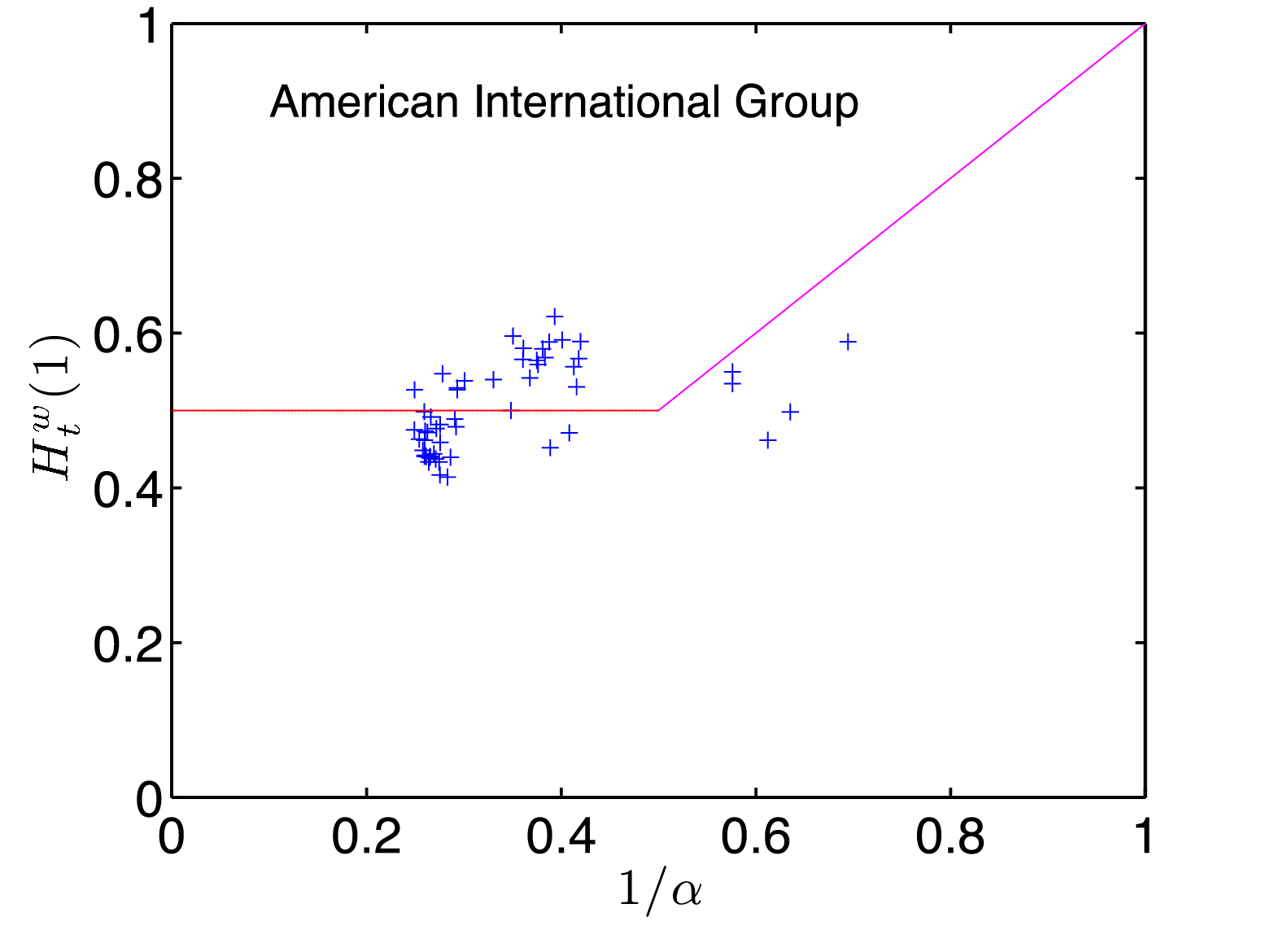} }\quad
 \subfigure{\includegraphics[width=0.5\columnwidth]{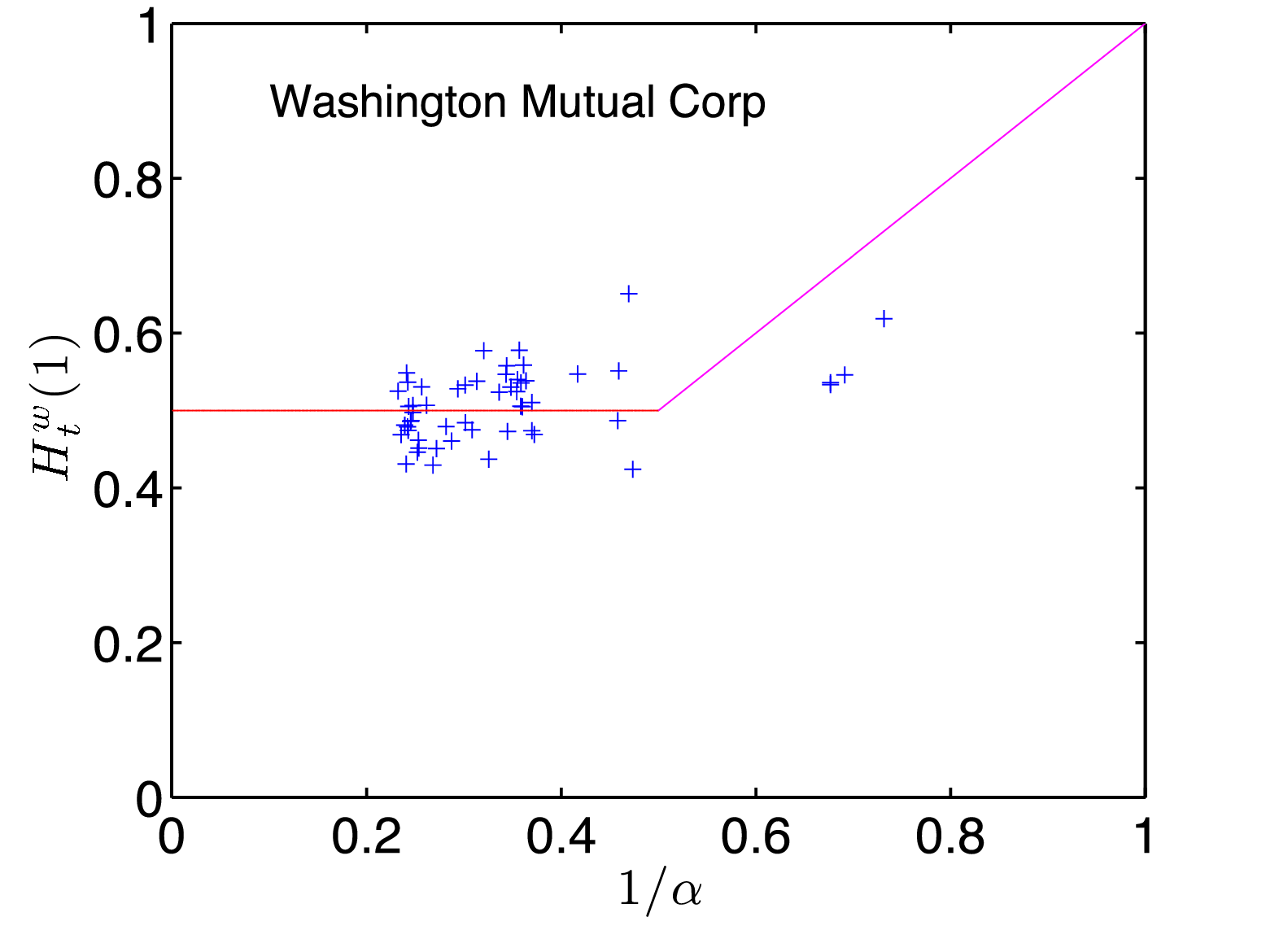} }} 
\mbox{\subfigure{\includegraphics[width=0.5\columnwidth]{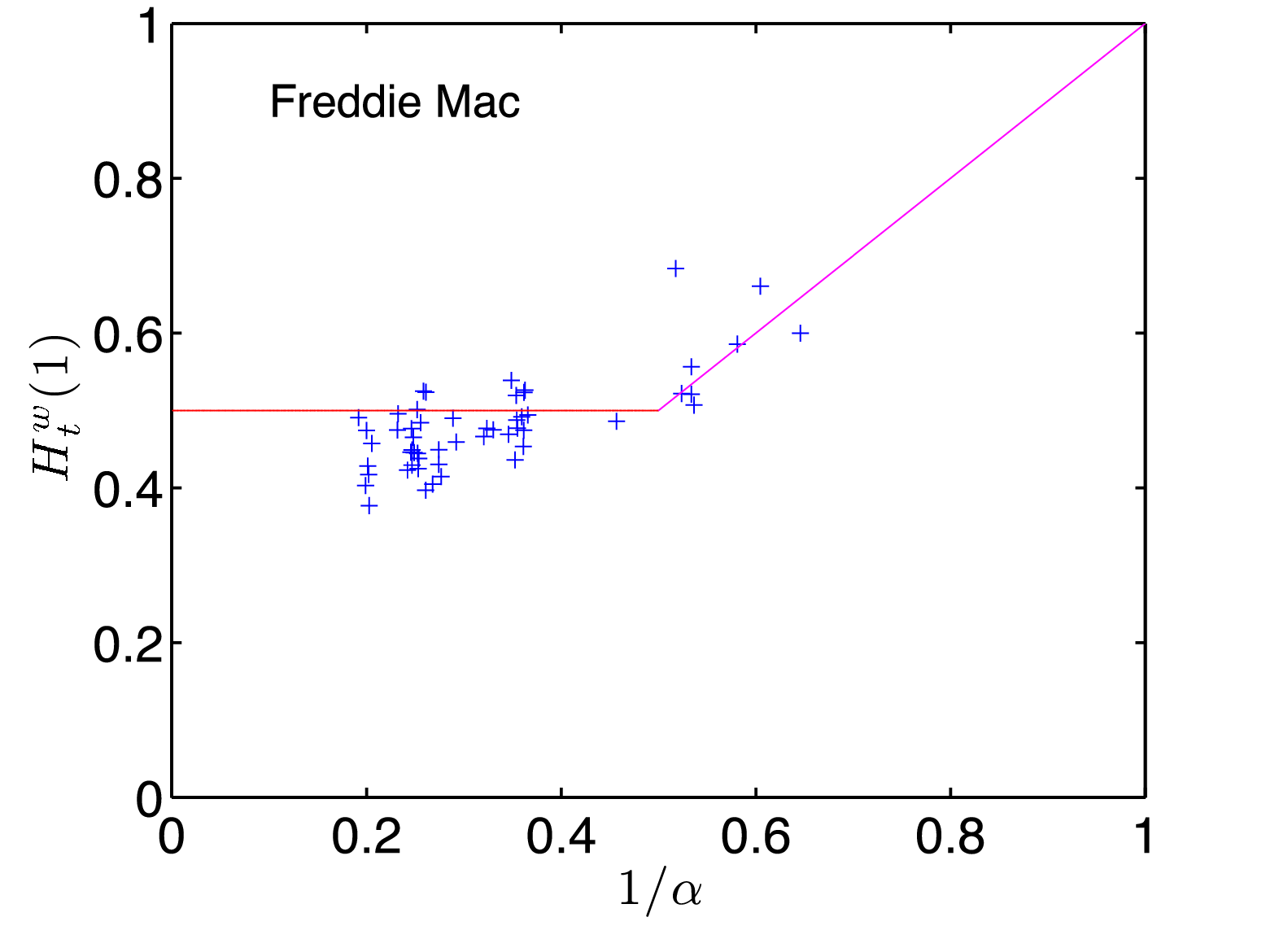} }\quad
 \subfigure{\includegraphics[width=0.5\columnwidth]{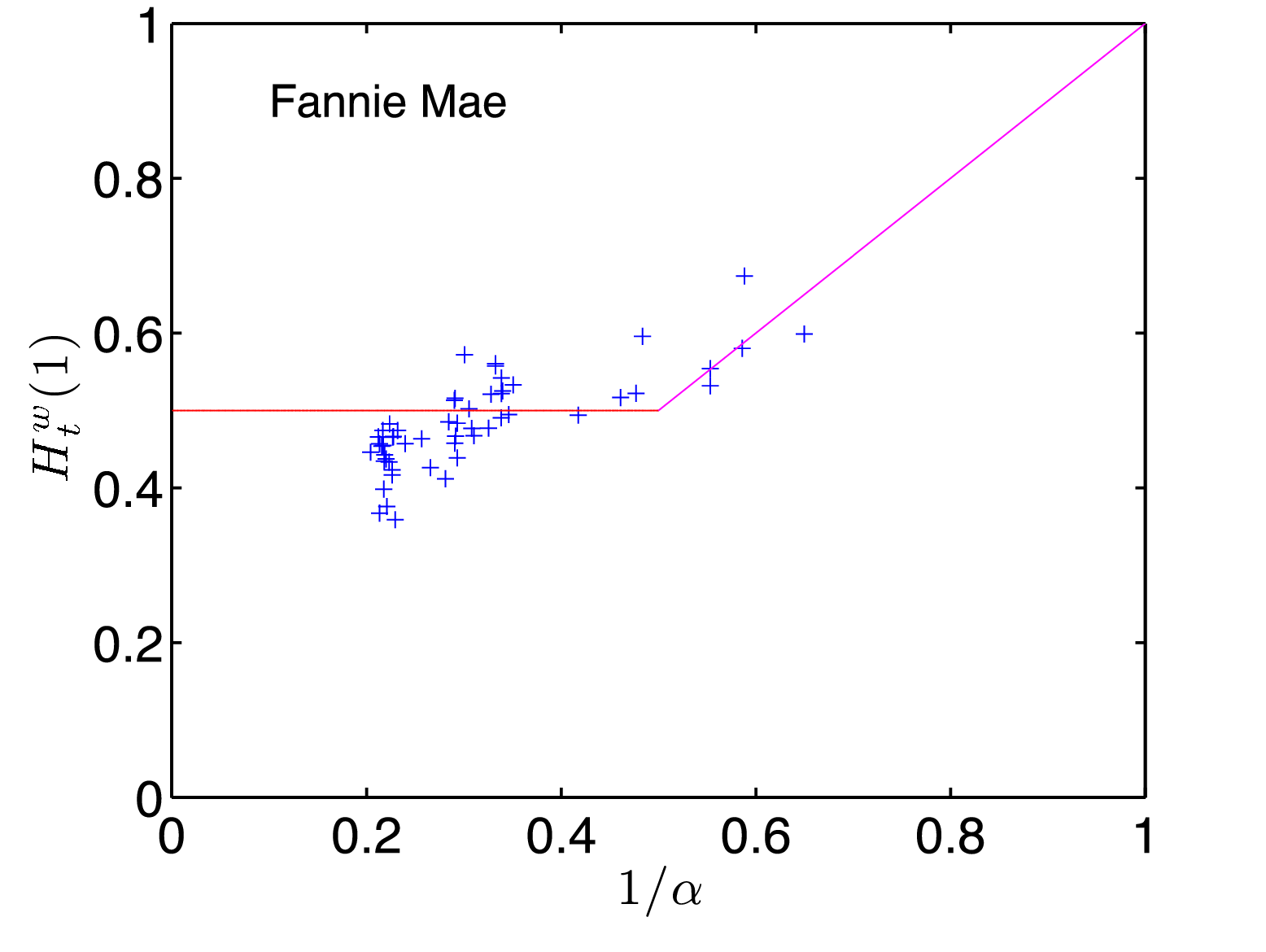} }}
\label{Figure10}
\caption{\textit{
Plots of the weighted Generalized Hurst exponent $H^w(1)$ versus the tail exponents for some bailed-out companies. 
From top left clockwise: American International Group, Washington Mutual Group, Freddie Mac and Fannie Mae. 
Both wGHE and tail exponents are computed using time-windows of $\Delta t = 750$ days. The weighted average for $H^w(1)$ is implemented with $\theta$=250 days.
}}
\end{figure}
By comparing Eq.$\ref{eq14}$ with Eq.$\ref{eq15}$ we get 
\begin{equation}\label{Hscal}
H=\left\{
\begin{array}{ll}
1/\alpha
& \mbox{if $\alpha < 2$}\\
\\
1/2
& \mbox{if $\alpha \ge 2$} \;. \\
\end{array} \right. 
\end{equation}
Eq.$\ref{eq15}$ holds also for the moments computed using the weighted average, by substituting $H$ with $H^w$ and the expectation values $E(\dots)$ with weighted averages. 
Processes with the property in Eq.$\ref{eq15}$ are deemed uniscaling. 
For $\alpha\ge 2$ we retrieve $H\sim0.5$ and the processes scales as a Brownian motion.
Let us here stress that the result in Eq.\ref{Hscal} is only valid for a random-walk type iid process with defined noise distribution.
On the other hand, it is well known that financial time series cannot be described within this framework. 
However, Eq.\ref{Hscal} is a valuable reference which can be used as a tool to compare the relation between the tail exponent and the Hurst exponent in more complex signals.

\begin{figure}[t!]
\centering
\mbox{\subfigure{\includegraphics[width=0.5\columnwidth]{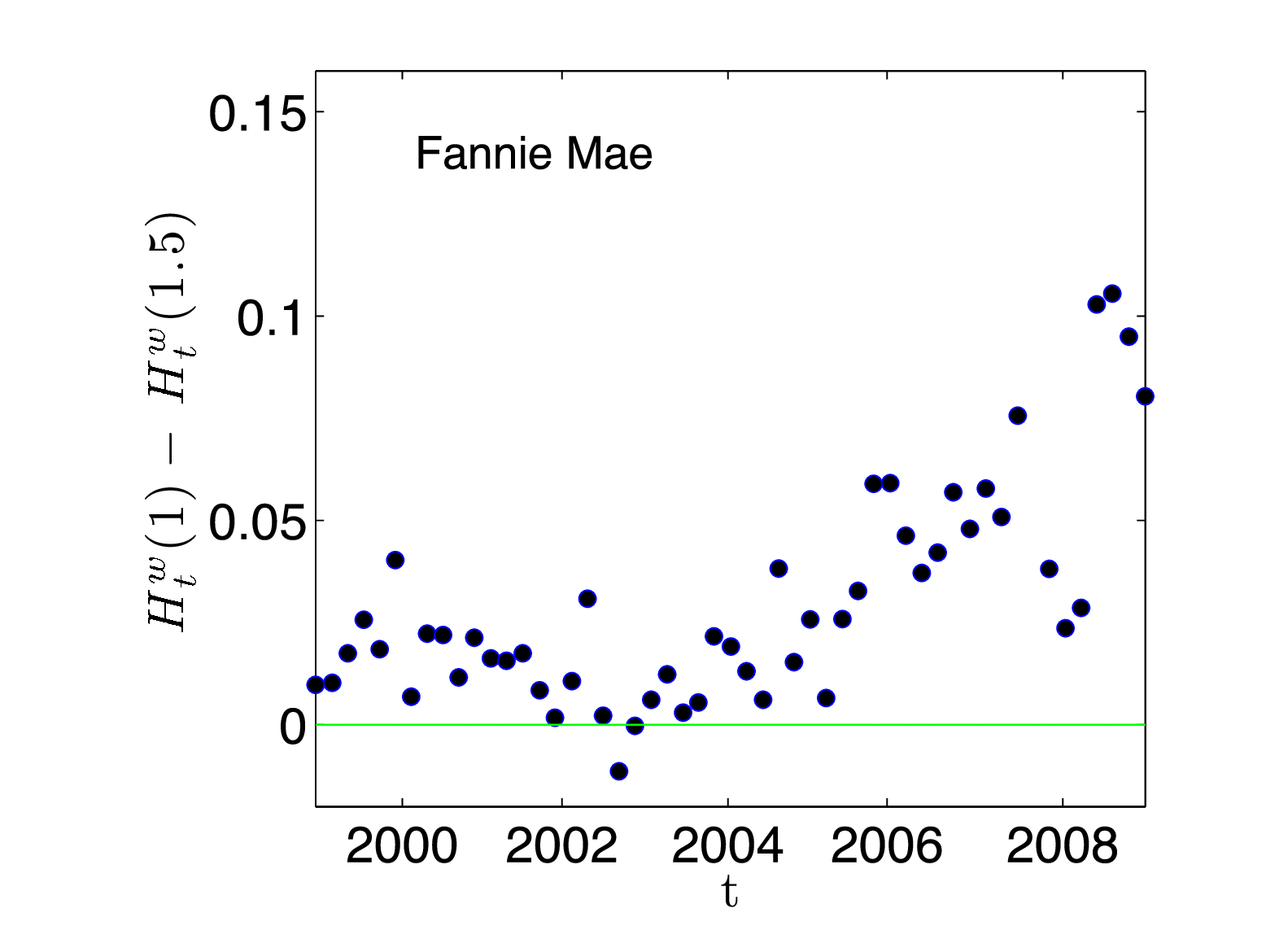}}\quad
\subfigure{\includegraphics[width=0.5\columnwidth]{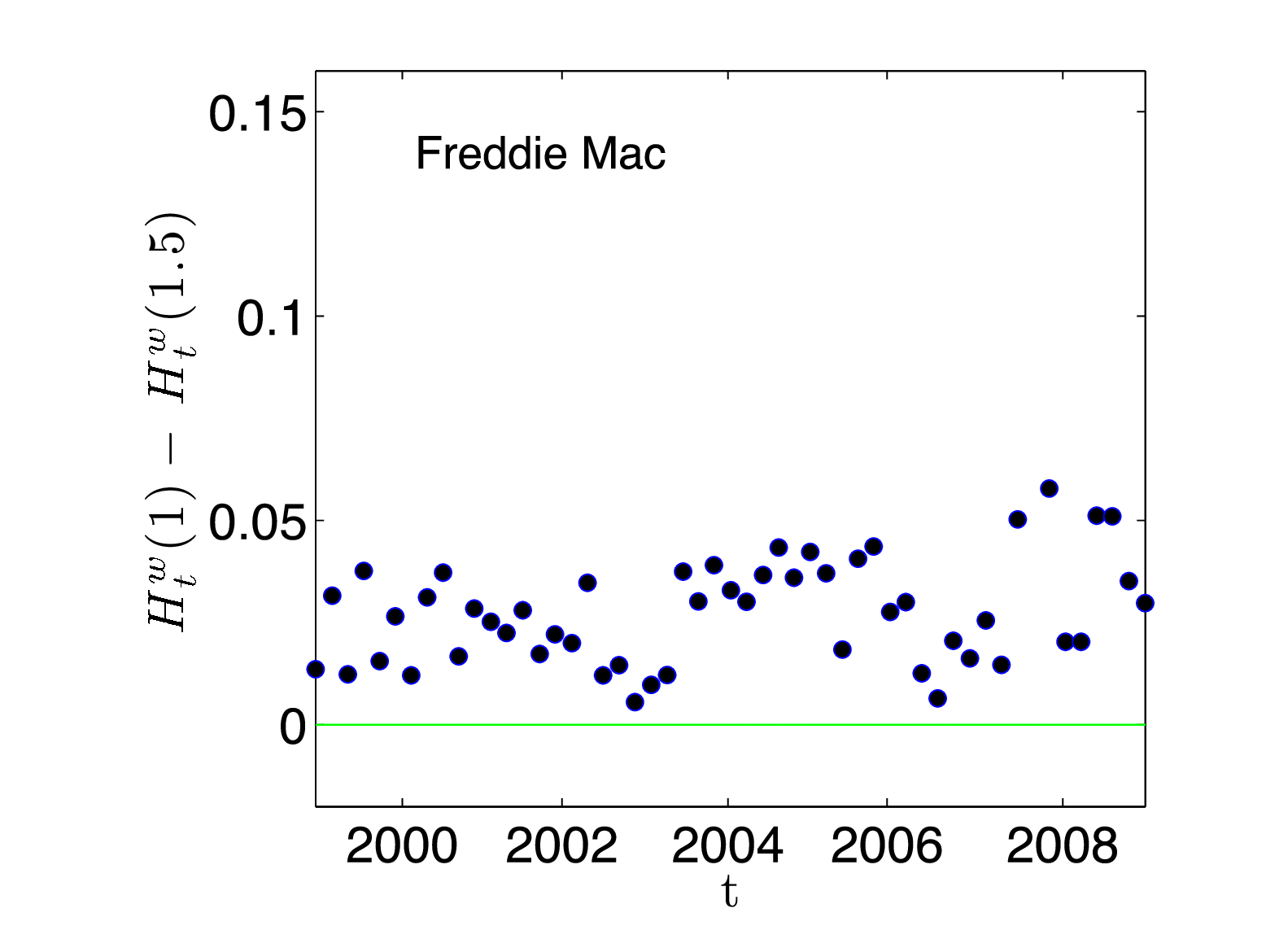} } }
\mbox{\subfigure{\includegraphics[width=0.5\columnwidth]{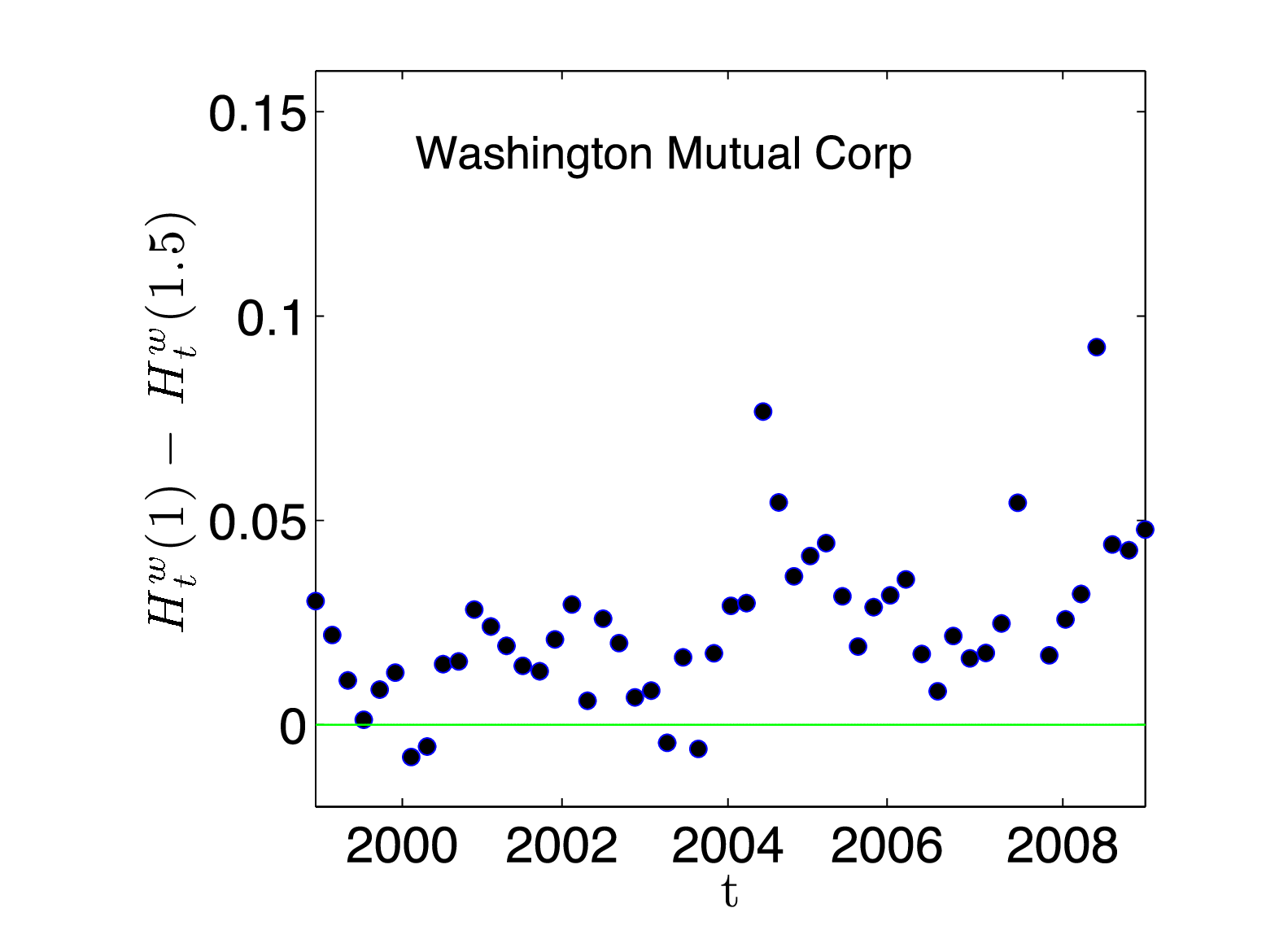}}\quad
\subfigure{\includegraphics[width=0.5\columnwidth]{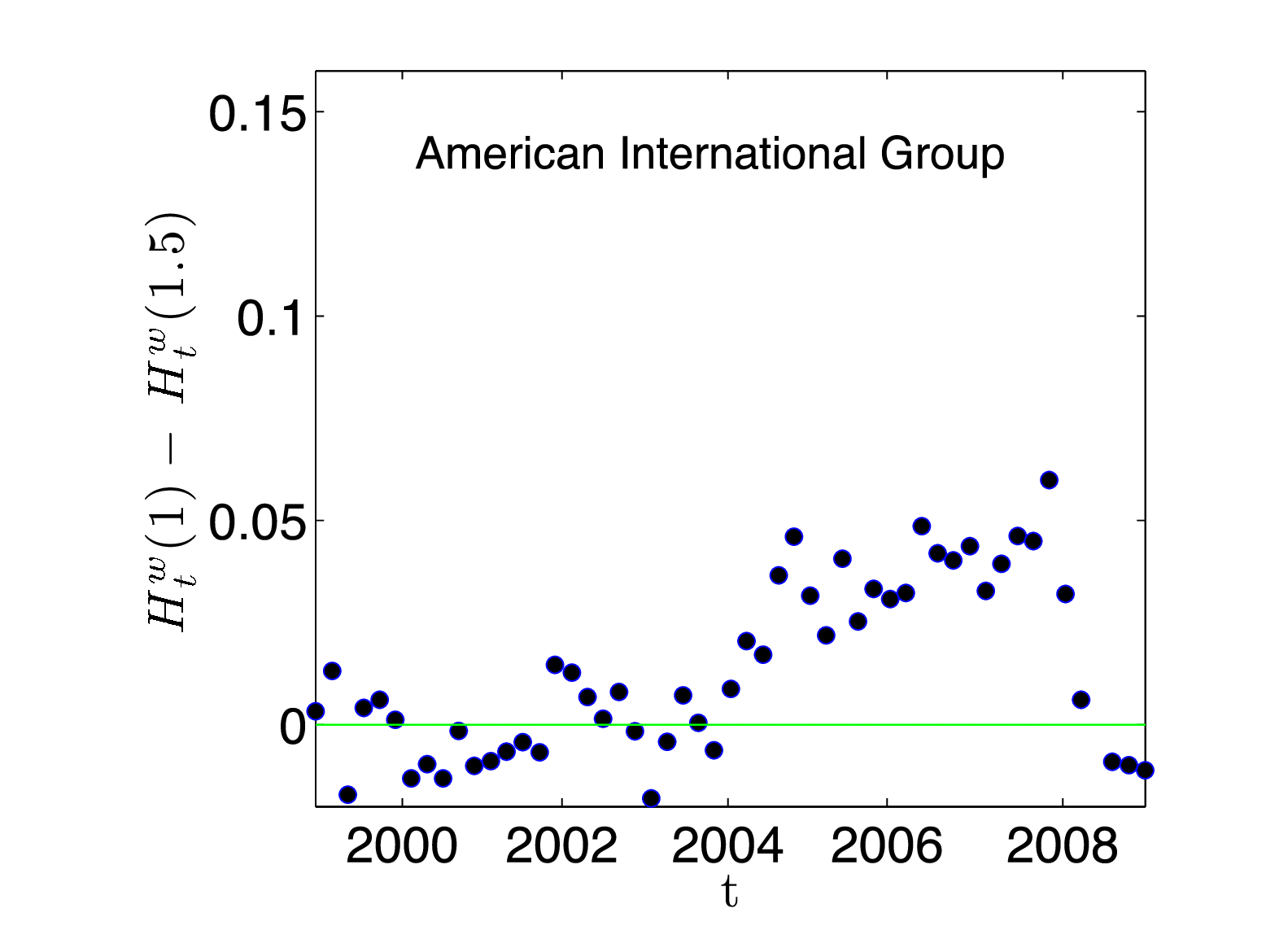} } }
\caption{\textit{
The difference $H^w(1)-H^w(1.5)$ as function of time. 
This quantity is a simple measure of multifractality of the system, as it quantifies the departure of the wGHE from the unifractal value $H\sim0.5$. From top left clockwise: Fannie Mae, Freddie Mac, Washington Mutual Group and American International Group. The parameters for the weighted mean are $\theta=250$ and $\Delta t=750$ days.}}
\label{Figure11}
\end{figure}

In Fig.10 we report plots of the wGHE, $H^w(1)$, versus the tail exponents computed at different times (see caption for details).
According to the previous considerations, we must observe a linear trend of $H^w(1)$ vs. $1/\alpha$ for $\alpha\le2$ and, instead, a flat behavior for $\alpha<2$.
In the figure we can indeed observe a departure from a linear trend occurring around $\alpha \sim 2$. 
On the other hand, the poor agreement with the prediction from Eq.\ref{Hscal} reveals that the process is not uniscaling and this is a signature of multifractality.
This multifractality can be measured  by tracking the difference $H^w(q)-H^w(q')$ over the time-windows (see Fig.$\ref{Figure11}$). 
Intriguingly, this difference remains stable for most of the time for all the companies reported in the figure but, instead, it increases as soon as the unstable period is reached, suggesting that the scaling properties of the time series change with the unfolding of the crisis. 
We stress that the behavior observed in the empirical data is not necessarily related to a change of the stochastic process underlying the financial time series. 
The increase in the multifractality of these kinds of signals is likely to occur in the presence of large price fluctuations. 
In this case indeed, the attitude of the investors, and thus the prices' movements, in the short period, are very rarely reflecting the price behavior over larger periods.   
\newline  Let us finally note that for the bailed-out companies it would also be interesting to look at $H^w(2)$, which, as we said, is associated to the scaling of the auto-correlation function of the time series. However, in spite of the behavior being very similar to that observed for $H^w(1)$, for $\alpha<2$ (which is the case for these companies) the second moment is not defined and thus it's difficult to interpret the real meaning of $H^w(2)$.
\section{Conclusions}
We have studied the scaling behavior in time of log-returns of the companies more severely affected by the `credit-crunch' crisis. 
The results obtained for these companies have been compared to those obtained for companies belonging to different market sectors, showing persistent differences. To allow a reasonable differentiation in the time series we have introduced a weighting procedure which renders recent events more significant that remote ones. 
With this exponential smoothing method we have computed the weighted Generalized Hurst exponent for overlapping time-windows spanning a period of 13 years (1996-2009). 
The bailed-out companies reveal an increasing trend which crosses 0.5 hinting therefore to a transition between different stages of development. 
This behavior, not observed for many other companies, including others belonging to the financial sector itself, might suggest that the wGHE is conveying important information about the stability of a company and that by tracking its value in time one could have a further tool to assess risk. 
A comparison with the scaling of the distributions of the log-returns shows that large fluctuation over a period may be related to the increase of the wGHE. 
We have also looked at the multifractal behavior in time of these companies revealing a multiscaling behavior with multifractality increasing when the crisis occurred. 
These empirical facts will be the basis of future work aiming to realistically model the price formation and evolution in financial markets \cite{dacorogna2001introduction,matteo2005long,liu2008multifractality,bartolozzi2004stochastic}. 
\bigskip 
\newline{\bf{Acknowledgements}}
\medskip
\newline We wish to thank the COST MP0801 project for partially supporting this work.
\bigskip
\bibliographystyle{unsrt}

\end{document}